\documentclass[fleqn, usenatbib]{mnras}
\usepackage{newtxtext,newtxmath}
\usepackage[T1]{fontenc}
\usepackage{ae,aecompl}
\usepackage{graphicx}
\usepackage{xcolor}
\usepackage{amsmath}
\usepackage{upgreek}
\usepackage{commath}
\usepackage{siunitx}
\usepackage[capitalise, noabbrev]{cleveref}
\usepackage{enumitem}
\usepackage{csvsimple}
\hypersetup{unicode}
\usepackage[utf8]{inputenc}

\DeclareRobustCommand{\VAN}[3]{#2}
\let\VANthebibliography\thebibliography
\def\thebibliography{\DeclareRobustCommand{\VAN}[3]{##3}\VANthebibliography}

\bibpunct[]{(}{)}{;}{a}{}{,}
\defcitealias{1981PASP...93....5B}{BPT}

\newcommand{\program}{\textsc}
\newcommand{\ssim}{\sim \!}
\newcommand{\Isample}{\citeauthor{2016Natur.529..178I} sample}

\newcommand{\lymana}{{Lyman-\ensuremath{\upalpha}}}
\newcommand{\lya}{{Ly\ensuremath{\upalpha}}}
\newcommand{\lyatext}{Lyα}
\newcommand{\HII}{{\ion{H}{II}}}
\newcommand{\CIV}{{\ion{C}{IV}}}
\newcommand{\HeII}{{\ion{He}{II}}}
\newcommand{\OIIIf}{{[\ion{O}{III}]}}
\newcommand{\OIIIs}{{\ion{O}{III}]}}
\newcommand{\CIII}{{\ion{C}{III}}}
\newcommand{\CIIIf}{{[\ion{C}{III}]}}
\newcommand{\CIIIs}{{\ion{C}{III}]}}
\newcommand{\MgII}{{\ion{Mg}{II}}}
\newcommand{\OII}{{[\ion{O}{II}]}}
\newcommand{\NeIII}{{[\ion{Ne}{III}]}}

\newcommand{\Ha}{{H\ensuremath{\upalpha}}}
\newcommand{\Hb}{{H\ensuremath{\upbeta}}}
\newcommand{\NII}{{[\ion{N}{II}]}}
\newcommand{\SII}{{[\ion{S}{II}]}}
\newcommand{\OI}{{[\ion{O}{I}]}}

\title[A spectroscopic study of a lensed galaxy at $z \sim 5$]{Assessing the sources of reionisation: a spectroscopic case study of a 30{$\times$} lensed galaxy at $z \sim 5$ with \texorpdfstring{\lya, \CIV, \MgII, and \NeIII}{\lyatext, CIV, MgII, and [NeIII]}}

\author[J. Witstok et al.]{{Joris Witstok,$^{1,2}$\thanks{E-mail: \href{mailto:jnw30@cam.ac.uk}{jnw30@cam.ac.uk}}
    Renske Smit,$^{1,2,3}$
    Roberto Maiolino,$^{1,2}$
    Mirko Curti,$^{1,2}$
    }
    \newauthor{Nicolas Laporte,$^{1,2}$
    Richard Massey,$^{4}$
    Johan Richard,$^{5}$
    and Mark Swinbank$^{4}$
    }
    \\
    $^{1}$Kavli Institute for Cosmology, University of Cambridge, Madingley Road, Cambridge CB3 0HA, UK
    \\
    $^{2}$Cavendish Laboratory, University of Cambridge, 19 JJ Thomson Avenue, Cambridge CB3 0HE, UK
    \\
    $^{3}$Astrophysics Research Institute, Liverpool John Moores University, 146 Brownlow Hill, Liverpool L3 5RF, UK
    \\
    $^{4}$Centre for Extragalactic Astronomy, Department of Physics, Durham University, South Road, Durham DH1 3LE, UK
    \\
    $^{5}$Univ Lyon, Univ Lyon1, Ens de Lyon, CNRS, Centre de Recherche Astrophysique de Lyon UMR5574, F-69230, Saint-Genis- Laval, France
}

\date{Accepted ---. Received ---; in original form ---}

\pubyear{2021}

\AtBeginDocument{
  \hypersetup{
    pdftitle={Preparing for the sources behind reionisation},
    pdfauthor={J. Witstok et al.},
    pdfsubject={A spectroscopic case study of a 30× lensed galaxy at z ∼ 5 with Lyα, CIV, MgII, and [NeIII]},
    pdfkeywords={{galaxies: high-redshift} -- {reionization} -- {gravitational lensing: strong} -- {methods: observational} -- {techniques: spectroscopic}}
  }
}

\begin{document}
\label{firstpage}
\pagerange{\pageref{firstpage}--\pageref{lastpage}}
\maketitle

\begin{abstract}
    We present a detailed spectroscopic analysis of a galaxy at $z\simeq4.88$ that is, by chance, magnified $\ssim30\times$ by gravitational lensing. Only three sources at $z\gtrsim5$ are known with such high magnification. This particular source has been shown to exhibit widespread, high equivalent width $\CIV\,\lambda\,1549\,\Angstrom$ emission, implying it is a unique example of a metal-poor galaxy with a hard radiation field, likely representing the galaxy population responsible for cosmic reionisation. Using UV nebular line ratio diagnostics, VLT/X-shooter observations rule out strong AGN activity, indicating a stellar origin of the hard radiation field instead. We present a new detection of $\NeIII\,\lambda\,3870\,\Angstrom$ and use the \NeIII/\OII\ line ratio to constrain the ionisation parameter and gas-phase metallicity. Closely related to the commonly used \OIIIf/\OII\ ratio, our \NeIII/\OII\ measurement shows this source is similar to local ``Green Pea'' galaxies and Lyman-continuum leakers. It furthermore suggests this galaxy is more metal poor than expected from the Fundamental Metallicity Relation, possibly as a consequence of excess gas accretion diluting the metallicity. Finally, we present the highest redshift detection of $\MgII\,\lambda\,2796\,\Angstrom$, observed at high equivalent width in emission, in contrast to more evolved systems predominantly exhibiting \MgII\ absorption. Strong \MgII\ emission has been observed in most $z\sim0$ Lyman-continuum leakers known and has recently been proposed as an indirect tracer of escaping ionising radiation. In conclusion, this strongly lensed galaxy, observed just $300\,\mathrm{Myr}$ after reionisation ends, enables testing of observational diagnostics proposed to constrain the physical properties of distant galaxies in the \textit{JWST}/ELT era.
\end{abstract}


\section{Introduction}
\label{sec:Introduction}

Space-based observatories such as \textit{Hubble Space Telescope} (\textit{HST}) and \textit{Spitzer} and ground-based $8 \, \mathrm{m}$-class telescopes have transformed our view of galaxy evolution in the high-redshift Universe, identifying statistically substantial samples of distant galaxies in deep imaging surveys beyond $z>4$ \citep{2014ARA&A..52..415M}. At this epoch, covering the first $\ssim 10\%$ of the current age of the Universe, the physical properties of galaxies were likely to be very different to those today, with metal-poor stellar populations, low stellar masses, and hard radiation fields. These conditions are favourable to strong nebular emission, despite the weak stellar continuum \citep[e.g.][]{2016ARA&A..54..761S}. Equally, this suggests the faint galaxy population in the Epoch of Reionisation (EoR) can contribute significantly to reionisation \citep[e.g.][]{2015ApJ...803...34B}.

This picture has mainly emerged from spectroscopic follow-up observations of individual distant sources selected in deep photometric surveys, although this poses several challenges. From the ground, near-infrared (NIR) spectrometers are restricted by Earth's atmosphere to observe key rest-frame optical emission line features, such as \Ha, $\OIIIf \, \lambda \, 5008 \, \Angstrom$, and $\OII \, \lambda \, 3727, 3730 \, \Angstrom$ (simply \OII\ hereafter) out to redshifts of about $2.5, 3.6$, and $5.2$, respectively. The much-anticipated \textit{James Webb Space Telescope} (\textit{JWST}) will explore the rest-frame optical spectra of more distant objects ($z \sim 4$-$12$), which will enable the use of many emission line diagnostics that are carefully calibrated with the wealth of data for more nearby galaxies, like the optical classification schemes that distinguish spectra of star-forming galaxies shaped by nebular emission from \HII\ regions from those dominated by emission of the narrow-line region of Active Galactic Nuclei (AGN; \citealt*{1981PASP...93....5B}, \citetalias{1981PASP...93....5B} hereafter; \citealt{1987ApJS...63..295V}).

Meanwhile, several new methods have been explored, which even as \textit{JWST} is launched will prove valuable in the era of Extremely Large Telescopes (i.e. ELT, GMT, and TMT). For example, alternative classification schemes to the \citetalias{1981PASP...93....5B} classification have been proposed, targeting the rest-frame ultraviolet (UV) instead: these use highly ionised gas lines such as $\CIIIs \, \lambda \, 1907 \, \Angstrom, \CIIIf \, \lambda \, 1909 \, \Angstrom$ (\CIII\ collectively), $\CIV \, \lambda \, 1548, 1551 \, \Angstrom$ (\CIV), and $\HeII \, \lambda \, 1640 \, \Angstrom$ (\HeII) to separate star-forming galaxies from AGN \citep[e.g.][]{2016MNRAS.456.3354F}. These lines are much brighter in the composite spectra of $z \sim 3$ Lyman-break galaxies than observed in the local Universe \citep[e.g.][]{2003ApJ...588...65S}. Another pressing challenge is to find a reliable method to uncover the sources responsible for reionisation by indirectly identifying Lyman-continuum (LyC) leakage in the EoR, where LyC and \ion{H}{I} \lymana\ (\lya) becomes inaccessible due to absorption by the neutral IGM. Methods aimed at characterising EoR galaxies, such as the UV classification schemes and indirect proxies of LyC escape, can be tested at (slightly) lower redshift where features are readily observable with current instrumentation, ideally with analogues of high-redshift galaxies.

In this work, we present one such case study, investigating in detail the emission line properties of RCS0224z5, a strongly lensed galaxy at redshift $z \simeq 4.88$ in the background of the RCS~0224--0002 cluster. RCS~0224--0002, a galaxy cluster at $z=0.773$, was discovered in the Red-Sequence Cluster Survey \citep[RCS;][]{2002AJ....123....1G}. Several arc-like structures were found in this study, among which an arc consisting of four images of a gravitationally lensed background galaxy at $z \simeq 4.88$, identified via its \lya\ emission. The magnification of the four images ranges from $\mu = 1.30$ to $\mu \sim 140$, making this only one of three known sources at $z \gtrsim 5$ with a comparably high magnification \citep{1997ApJ...486L..75F, 2009MNRAS.400.1121S, 2021ApJ...906..107K} -- and placing this galaxy at less than $300 \, \mathrm{Myr}$ after the end of reionisation. Follow-up observations with VLT/MUSE have furthermore revealed spatially widespread and narrow ($\text{FWHM} \simeq 156 \, \mathrm{km/s}$) \CIV\ emission with a high equivalent width (EW) of $\ssim 10 \, \Angstrom$ in the rest frame \citep{2017MNRAS.467.3306S}, similar to what is being observed in an increasing number of $z \sim 6$-$8$ galaxies \citep{2015MNRAS.454.1393S, 2017ApJ...836L..14M, 2017ApJ...851...40L}, but rarely seen in the local Universe \citep{2019ApJ...874...93B, 2019MNRAS.488.3492S}.

We present new VLT/X-shooter observations that constrain the rest-frame UV emission line diagnostics that are inaccessible within the MUSE wavelength range. Unlike sources at higher redshift, where no rest-frame optical features are accessible from the ground, \citet{2007MNRAS.376..479S} presented widespread \OII\ detected in deep SINFONI observations. We present the additional detection of $\NeIII \, \lambda \, 3870 \, \Angstrom$ emission and corresponding new measurement of the \NeIII/\OII\ line diagnostic to place this system in the context of the local galaxy population, in order to gain insight into the origin of high-EW \CIV\ emission in the early Universe. Finally, we report the detection of $\MgII \, \lambda \, 2796 \, \Angstrom$ in emission: a remarkable finding, as this is in stark contrast with the local galaxy population, where it is mostly observed in absorption \citep[e.g.][]{1993ApJS...86....5K}. Being a resonant transition like \lya, it has the potential to be an indirect tracer of LyC escape \citep[e.g.][]{2018ApJ...855...96H}.

The outline of this paper is as follows. In \cref{sec:Observations}, we describe the observations, and in \cref{sec:Results} we present the results. In \cref{sec:Discussion} we discuss the outcomes, and we finally summarise our findings in \cref{sec:Summary}. In our analysis, we adopt the cosmological parameters $\Omega_\text{m} = 0.3$, $\Omega_\Lambda = 0.7$, and $H_0 = 70 \, \mathrm{km \, s^{-1} \, Mpc^{-1}}$ throughout (implying an angular scale of $6.4 \, \mathrm{kpc/arcsec}$ at $z=4.88$), to ease comparison with previous studies. All magnitudes are in the AB system \citep{1983ApJ...266..713O}.

\begin{figure*}
	\centering
	\includegraphics[width=\linewidth]{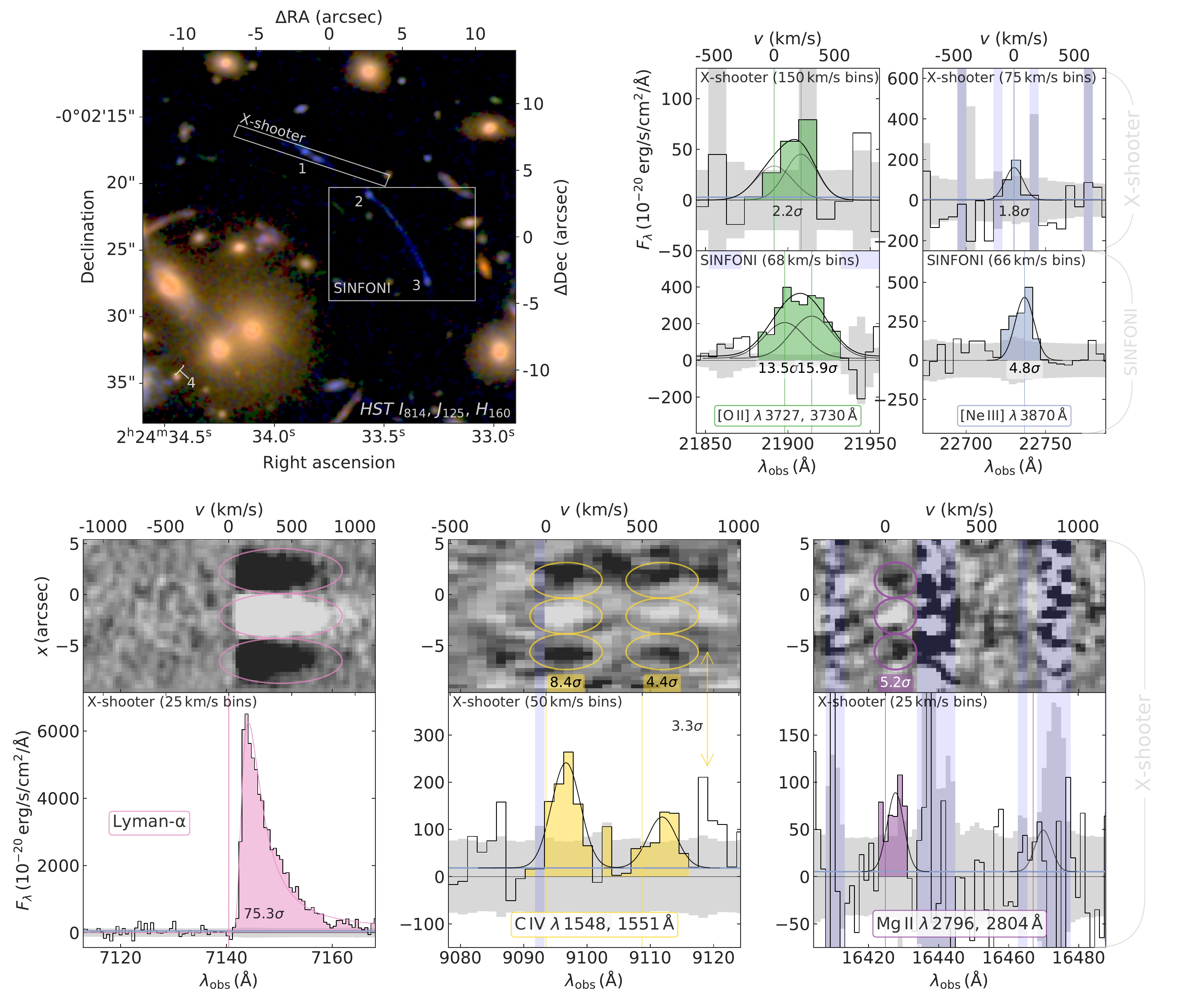}
	\caption[Overview of the observations discussed in \cref{ssec:Observations: X-shooter,ssec:Observations: SINFONI,ssec:Observations: HST}.]{Overview of the observations discussed in \cref{ssec:Observations: X-shooter,ssec:Observations: SINFONI,ssec:Observations: HST}. \textit{Top left}: \textit{HST} false-colour $I_{814}$, $J_{125}$, and $H_{160}$ image of RCS~0224--0002, indicating the lensed galaxy images 1$-$4 and the FOV of the observations covering the arc of RCS0224z5. \textit{Top right}: \OII\ and \NeIII\ in both X-shooter and SINFONI spectra. \textit{Bottom row}: spectra of \lya, \CIV, and \MgII\ emission lines observed with VLT/X-shooter. An arrow marks a tentative $\ssim 3 \sigma$ redshifted component of \CIV, see \cref{ssec:Results: X-shooter}. In two-dimensional X-shooter spectra, the dark-light-dark signature of detected lines are highlighted with coloured ellipses (\cref{ssec:Observations: X-shooter}). Labels indicate the significance of each detection measured in a SNR-optimised aperture (larger for \lya, smaller for \CIV\ and \MgII), while all one-dimensional spectra shown are extracted from a larger aperture to capture the entire flux as reported in \cref{tab:Results} (see \cref{ssec:Results: X-shooter} for details). The grey filled-in area shows the $1 \sigma$ uncertainty level. The rest-frame UV continuum fit is shown with a blue line. Velocities are based on the corresponding systemic redshift of \OII\ (\cref{sec:Results}) and are centred on the brightest line for doublets.
	}
	\label{fig:Overview panel}
\end{figure*}

\section{Observations}
\label{sec:Observations}

\begin{table*}
    \centering
    \caption[Observational results]
    {Overview of the observed emission lines in the X-shooter and SINFONI spectra. For non-detections, upper limits are given (see \cref{ssec:Results: X-shooter} for details).
    }
    \begin{tabular}{lcccccc}
        \hline
        Instrument & Image(s)$^a$ & Feature & $\lambda_\text{vac} \, (\Angstrom)^b$ & $\Delta v \, (\mathrm{km/s})^c$ & $\mathrm{Flux} \, (10^{-18} \, \mathrm{erg \, s^{-1} \, cm^{-2}})^d$ & $\mathrm{EW} \, (\Angstrom)^e$
        \\
        \hline
        \begin{filecontents*}{XResults.csv}
eline,wl,dv,dverr,uplim,flux,fluxerr,EW,EWerr,FWHM,FWHMerr
\lya,1215.67,156,52,False,440.4,5.9,143.3,12.9,207,6
\CIV,1548.19,104,53,False,12.7,3.1,11.9,3.2,171,30
\CIV,1550.77,104,53,False,6.18,3.48,5.78,3.34,171,30
\HeII,1640.42,nan,nan,True,6.11,nan,6.21,nan,nan,nan
\CIIIf,1906.68,nan,nan,True,26.7,nan,35.2,nan,nan,nan
\CIIIs,1908.73,nan,nan,True,46.0\,^f,nan,25.4\,^f,nan,nan,nan
\MgII,2796.35,52,54,False,5.00,1.49,16.2,5.1,87,46
\end{filecontents*}
\csvreader[late after line=\\, head to column names]{XResults.csv}{}{\csviffirstrow{\textit{X-shooter} & 1 &}{&&}
            \eline & \wl & \ifcsvstrcmp{\dv}{nan}{\dots}{$\dv \ifcsvstrcmp{\dverr}{nan}{}{\pm \dverr}$} & \ifcsvstrcmp{\flux}{nan}{\dots}{\ifcsvstrcmp{\uplim}{True}{$<\flux$}{$\flux \pm \fluxerr$}} & \ifcsvstrcmp{\EW}{nan}{\dots}{\ifcsvstrcmp{\uplim}{True}{$<\EW$}{$\EW \pm \EWerr$}}
        }
        \hline
        \begin{filecontents*}{SResults.csv}
eline,wl,dv,dverr,uplim,flux,fluxerr,EW,EWerr,FWHM,FWHMerr
\OII,3727.09,0,13,False,60.5,4.5,nan,nan,355,71
\OII,3729.88,0,13,False,71.4,4.5,nan,nan,355,71
\NeIII,3869.85,0,13,False,60.4,12.6,nan,nan,55,141
\end{filecontents*}
\csvreader[late after line=\\, head to column names]{SResults.csv}{}{\csviffirstrow{\textit{SINFONI} & 2 and 3 &}{&&}
            \eline & \wl & \ifcsvstrcmp{\dv}{nan}{\dots}{$\dv \ifcsvstrcmp{\dverr}{nan}{}{\pm \dverr}$} & \ifcsvstrcmp{\flux}{nan}{\dots}{\ifcsvstrcmp{\uplim}{True}{$<\flux$}{$\flux \pm \fluxerr$}} & \ifcsvstrcmp{\EW}{nan}{\dots}{\ifcsvstrcmp{\uplim}{True}{$<\EW$}{$\EW \pm \EWerr$}}
        }
    \end{tabular}
    \flushleft
    $^a$ Lensed image(s) from which the spectroscopic measurements were taken (see \cref{ssec:Observations: X-shooter,ssec:Observations: SINFONI} for details).
    \\
    $^b$ Vacuum rest-frame wavelength.
    \\
    $^c$ Velocity offset with respect to the systemic redshift as measured by the relevant instrument (X-shooter: $z_\text{sys} = 4.8737$, SINFONI: $z_\text{sys} = 4.8754$). Uncertainty in the velocity offsets includes the uncertainty in determining the systemic redshift (see \cref{sec:Results}).
    \\
    $^d$ Observed fluxes, uncorrected for the lensing magnification factor of $\mu = 29_{-11}^{+9}$ of image 1, observed by X-shooter (\cref{ssec:Observations: X-shooter}), and $\mu = 21_{-8}^{+12}$ and $\mu = 138_{-74}^{+7}$ of image 2 and 3 respectively, observed collectively by SINFONI (\cref{ssec:Observations: SINFONI}).
    \\
    $^e$ Rest-frame equivalent width (EW; positive values indicating a feature is observed in emission). Note that EWs are independent of lensing magnification.
    \\
    $^f$ Upper limits inferred under the assumption of a minimum ratio of $F_{1907}/F_{1909} \approx 1.39$ for $n_e \leq 10^3 \, \mathrm{cm^{-3}}$ (see \cref{ssec:Results: X-shooter}).
    \label{tab:Results}
\end{table*}

\subsection{X-shooter spectroscopy}
\label{ssec:Observations: X-shooter}

Observations of the lensed image 1 of RCS0224z5, amplified by $\mu = 29_{-11}^{+9}$ \citep[luminosity-weighted; see][]{2017MNRAS.467.3306S}, were taken with the VLT/X-shooter \citep{2011A&A...536A.105V} on 11, 14, and 16 October 2018 with a total on-source time of $3.5 \, \mathrm{h}$, under ESO programme ID 0102.A-0704(A) (PI: Smit); the slit was centred at $\alpha = 02$:24:33.83, $\delta = -00$:02:17.91 (\cref{fig:Overview panel}) and using slit widths of $1.2 \arcsec$ and $0.9 \arcsec$ in the visible (VIS) and near infrared (NIR), resulting in a spectral resolution $R \equiv \lambda/\Delta \lambda \approx 6500$ and $R \approx 5600$, respectively, in the two arms. Observations were taken with AB nodding with an offset of $3.8 \arcsec$ and individual exposures were $383 \, \mathrm{s}$ in the VIS and $230 \, \mathrm{s}$ NIR arm. Observations were taken with an average airmass of $1.14$ and seeing of $0.8 \arcsec$. Data reduction was performed using the standard \program{esoreflex} X-shooter pipeline \citep{2013A&A...559A..96F}. We apply the nodding-mode reduction, as well as stare-mode reduction with a manual algorithm to combine frames from the `ABBA' nodding pattern, depending on which yields the best results; the manual stare-mode reduction is used throughout, except for the \CIV\ line. Individual OBs were separately corrected for telluric absorption in the VIS and NIR arms using \program{molecfit} \citep{2015A&A...576A..77S, 2015A&A...576A..78K}.

\subsection{SINFONI spectroscopy}
\label{ssec:Observations: SINFONI}

Reduction of the SINFONI IFU spectroscopy is described in \citet{2007MNRAS.376..479S}. In short, IFU spectroscopy was performed for a total of $12 \, \mathrm{h}$ on source with VLT/SINFONI \citep{2003SPIE.4841.1548E, 2004Msngr.117...17B} under ESO programme ID 075.B-0636(B). The data were taken in the HK grating, resulting in a spectral resolution of $R \approx 1700$, with a $\ssim 8 \times 8 \, \mathrm{arcsec^2}$ field of view (at a spatial sampling of $0.25 \, \mathrm{arcsec/pixel}$), covering the lensed images 2 and 3 of RCS0224z5 (luminosity-weighted amplifications of $\mu = 21_{-8}^{+12}$ and $\mu = 138_{-74}^{+7}$, respectively; \citealt{2017MNRAS.467.3306S}). Lensed image 3 has a particularly high amplification (and corresponding uncertainty), as the arc crosses the lensing critical curve. We note that the source plane image of the galaxy is fully recovered by lensed image 1, but only partially in images 2 and 3: out of the two clumps seen in image 1, the one in north east is not reproduced in 2 and 3 \citep[see][ for details]{2017MNRAS.467.3306S}.

\subsection{\textit{HST} imaging}
\label{ssec:Observations: HST}

\textit{HST} imaging of RCS~0224--0002 is available on the Space Telescope Science Institute data archive\footnote{Data may be obtained from the MAST at \href{https://dx.doi.org/10.17909/t9-9kg5-hg27}{10.17909/T9-9KG5-HG27}.} \citep[GO 14497, PI: Smit; see][]{2017MNRAS.467.3306S}. Observations were performed with the Advanced Camera for Surveys (ACS) using the F814W ($I_{814}$) filter ($2.2 \, \mathrm{ks}$ exposure), and with the Wide Field Camera 3 (WFC3) using the F125W ($J_{125}$) and F160W ($H_{160}$) filters (both $2.6 \, \mathrm{ks}$ exposures). The resulting images reach a depth of $26.3 \, \mathrm{mag}$, $26.8 \, \mathrm{mag}$, and $26.7 \, \mathrm{mag}$ in the $I_{814}$, $J_{125}$, and $H_{160}$ bands ($5\sigma$ for a $0.5 \arcsec$-diameter aperture). A false-colour image of these three bands is shown in the bottom left of \cref{fig:Overview panel}.

\section{Results}
\label{sec:Results}

In this study, we are mainly interested in line diagnostics using \lya, \CIV, \HeII, the \CIII\ doublet, $\MgII \, \lambda \, 2796, 2804 \, \Angstrom$ (\MgII), \OII, and $\NeIII \, \lambda \, 3870 \, \Angstrom$ (\NeIII).\footnote{In this work, we use vacuum wavelengths throughout (see \cref{tab:Results}); emission line labels reflect vacuum wavelengths, rounded to the nearest integer.} The measured velocity offsets, line fluxes, and rest-frame EWs (or upper limits) of these lines presented in \cref{fig:Overview panel} are summarised in \cref{tab:Results}. In the following paragraphs, we will discuss the results for the X-shooter and SINFONI data sets individually.

We derive the systemic redshift from the $\OII \, \lambda \, 3727, 3730 \, \Angstrom$ doublet in combination with the $\NeIII \, \lambda \, 3870 \, \Angstrom$ line for increased precision; for X-shooter, this is measured to be $z_\text{sys} = 4.8737 \pm 0.0010$ (uncorrected for a negligible barycentric velocity of $5$-$7 \, \mathrm{km/s}$), while for SINFONI it is $z_\text{sys} = 4.8754 \pm 0.0003$ (again not corrected for a barycentric velocity of up to $27 \, \mathrm{km/s}$ or $\Delta z = 0.0005$, but consistent with $z_\OII = 4.8757 \pm 0.0005$ measured by \citealt{2007MNRAS.376..479S}), leaving a difference of $\Delta z = 0.0017$. In both cases, we simultaneously fit the \OII\ and \NeIII\ lines while fixing the \OII\ line ratio, due to limited spectral resolution and signal to noise. We adopt a ratio of $F_{3730}/F_{3727} = 1.18$, the median of $z \sim 2.3$ star-forming galaxies reported by \citet{2016ApJ...816...23S}, corresponding to an electron density $\ssim 10^2 \, \mathrm{cm^{-3}}$. Importantly, the \OII\ flux, and hence the \NeIII/\OII\ line ratio, are practically invariant when fitting with a freely varying \OII\ line ratio. For X-shooter, the fit was performed on spectra rebinned to $150 \, \mathrm{km/s}$ and $75 \, \mathrm{km/s}$ for \OII\ and \NeIII\ respectively, using the intrinsic line widths obtained from the fit to the SINFONI spectrum. The $\ssim 2 \sigma$ discrepancy between the systemic redshifts determined by X-shooter and SINFONI, equivalent to $88 \pm 52 \, \mathrm{km/s}$, may be a calibration problem (exceeding barycentric velocity corrections); or, given that the two spectra probe different images (see \cref{fig:Overview panel}), the velocity difference can be a consequence of real kinematic differences between the two components. In the following, we measure offsets from the systemic redshift using the consistent measurement in the same image, and by the same instrument. We include the uncertainty in determining the systemic redshift (i.e. $52 \, \mathrm{km/s}$ for X-shooter) in the uncertainty on all velocity offsets.

For all line flux measurements, we fit Gaussian profiles to the one-dimensional spectra. The uncertainty is estimated by scaling the flux uncertainty of a single spectral channel, $F_\lambda \, \Delta \lambda$, by the square root of the number of spectral bins where line flux is detected (coloured channels in \cref{fig:Overview panel}).

\subsection{X-shooter}
\label{ssec:Results: X-shooter}

The relevant X-shooter data are presented in the bottom row of \cref{fig:Overview panel}, which shows one-dimensional spectra extracted from a $2.4 \arcsec$ aperture for all lines except \lya, where we use a $3.2 \arcsec$ aperture. These spectra are used to measure the velocity offsets and total fluxes (\cref{tab:Results}). The annotated labels show the signal-to-noise ratio (SNR) measured in a smaller $1.4 \arcsec$ aperture, again except for \lya\ where the extended aperture yields a higher SNR. We detect strong \lya\ emission (at $\ssim 80 \sigma$ in a $3.2 \arcsec$ aperture or $\ssim 50 \sigma$ in a $1.4 \arcsec$ aperture; in the $1.4 \arcsec$ aperture, the \lya\ EW decreases to $78 \, \Angstrom$), as well as the \CIV\ doublet, at $8.4 \sigma$ and $4.4 \sigma$. Finally, we detect the $\MgII \, \lambda \, 2796 \, \Angstrom$ line at $5.2 \sigma$ (see \cref{ap:MgII and CIII significance} for more details on the significance of this detection). This makes RCS0224z5 the highest redshift galaxy for which \MgII\ emission has been detected. After \lya, the $\MgII \, \lambda \, 2796 \, \Angstrom$ line has the highest EW out of all emission features detected in the rest-frame UV, even though the other line of the doublet, $\MgII \, \lambda \, 2804 \, \Angstrom$, is undetected due to skylines. We show the expected signal for $\MgII \, \lambda \, 2804 \, \Angstrom$ in \cref{fig:Overview panel} assuming a typical flux ratio of $F_{2796}/F_{2804} \approx 1.9$ between the $\MgII$ lines at $2796 \, \Angstrom$ and $2804 \, \Angstrom$ \citep[e.g.][]{2018ApJ...855...96H}, which is indeed below the estimated uncertainty level.

The blue line in the \CIV\ doublet, $\CIV \, \lambda \, 1551 \, \Angstrom$, appears to have a weak second redshifted component ($\ssim 3 \sigma$), although the negative signature in the two-dimensional spectrum seems mostly absent, which is why we do not include it in our analysis. At the current sensitivity and spectral resolution, we cannot confidently explain the nature of this feature; however, note that if this emission feature is included in the \CIV\ flux and EW, this would not affect our conclusions in \cref{ssec:Discussion: CIV} regarding the origin of highly ionised emission.

With X-shooter, we detect the \OII\ doublet and \NeIII\ at $2.2 \sigma$ and $1.8 \sigma$ by rebinning to $150 \, \mathrm{km/s}$ and $75 \, \mathrm{km/s}$, respectively (both in a $1.4 \arcsec$ aperture to maximise SNR). In our analysis, however, we will adopt the measurements of SINFONI at higher significance (see \cref{ssec:Results: SINFONI}). Furthermore, as shown in \cref{tab:Results}, we obtain upper limits on emission from the \HeII\ line and the \CIII\ doublet. Since both lines of the \CIII\ doublet fall directly on skylines, instead of a $2\sigma$ limit from the noise as for \HeII, we take the integrated flux measured within $-100 \, \mathrm{km/s} < v < 100 \, \mathrm{km/s}$ of the expected line centre of the $1907 \, \Angstrom$ line (which is slightly less affected by telluric absorption and skylines; see also \cref{ap:MgII and CIII significance}). We obtain an upper limit for the total flux of the doublet using the lowest physically attainable value of $F_{1907}/F_{1909} \approx 1.39$ for $n_e \leq 10^3 \, \mathrm{cm^{-3}}$ \citep{2019ApJ...880...16K}. If we take a ratio of $\ssim 0.34$ for $n_e \leq 10^5 \, \mathrm{cm^{-3}}$, the resulting \CIV/\CIII\ ratio we measure shifts by $0.36 \, \mathrm{dex}$, leaving our findings unaffected (\cref{ssec:Discussion: CIV}). The $\OIIIs \, \lambda \, 1661, 1666 \, \Angstrom$ doublet also remains undetected: the brightest line of the two, at $1666 \, \Angstrom$, falls on a skyline, and the second line at $1661 \, \Angstrom$ is too faint to provide useful upper limits on the doublet.

By rebinning to a lower spectral resolution (in bins of $\Delta \lambda_\text{obs} = 200 \, \Angstrom$, masking skylines prior to rebinning), we detect the rest-frame UV continuum and assuming $F_\lambda \propto \lambda^\beta$, we measure a UV-continuum slope $\beta = -2.36 \pm 0.28$, in good agreement with $\beta = -2.19 \pm 0.14$ as measured from the \textit{HST} $J_{125} - H_{160}$ colour \citep{2017MNRAS.467.3306S}. This continuum fit is shown in all X-shooter spectra in \cref{fig:Overview panel} with a blue line. Using this fit, we deduce the equivalent widths (or upper limits thereof) of observed lines.

\subsection{SINFONI}
\label{ssec:Results: SINFONI}

In addition to the previously reported observation of the \OII\ doublet with SINFONI \citep[\OII;][]{2007MNRAS.376..479S}, at $13.5 \sigma$ and $15.9 \sigma$ respectively, we present a new $4.8 \sigma$ detection of the \NeIII\ line (see \cref{fig:Overview panel}), the highest redshift detection of this line to date. The \NeIII\ feature is not confidently detected with X-shooter, likely due to the shorter exposure time ($3.5 \, \mathrm{h}$ versus $12 \, \mathrm{h}$).

Conversely, even though the SINFONI observations cover the wavelength of \MgII, it was not detected due to the lower spectral resolution of SINFONI ($R \approx 1700$ or $\Delta \lambda_\text{obs} \approx 10 \, \Angstrom$ at the observed wavelength of \MgII, $16426 \, \Angstrom$)\ blending the signal with the strong skyline feature at $16435 \, \Angstrom$.

\section{Discussion}
\label{sec:Discussion}

\begin{figure}
	\centering
	\includegraphics[width=\linewidth]{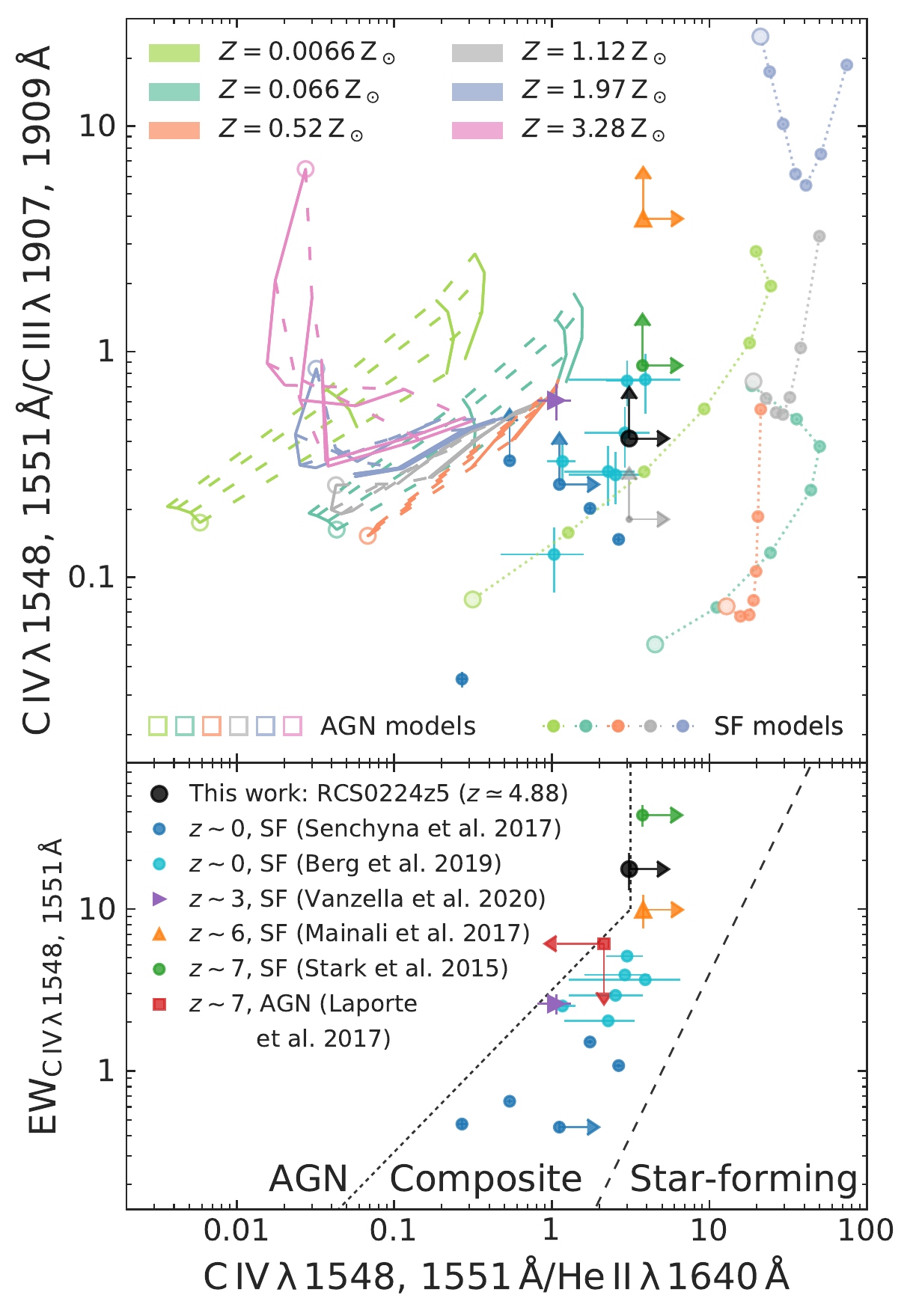}
	\caption[\CIV/\HeII line ratio diagnostics.]{\textit{Top panel}: line ratio of \CIV/\CIII\ versus \CIV/\HeII. The photoionisation models show predictions for nebular emission \citep[indicated as star-forming or SF; models from][]{2016MNRAS.462.1757G} and narrow-line region AGN emission \citep[models from][]{2016MNRAS.456.3354F}; for both, see the description in the text for details. The black and grey measurements show upper limits on the line ratios for RCS0224z5 assuming a maximum electron density of $n_e \leq 10^3 \, \mathrm{cm^{-3}}$ and $10^5 \, \mathrm{cm^{-3}}$ respectively (shifting the ratio by $0.36 \, \mathrm{dex}$, see \cref{ssec:Results: X-shooter}), both strongly rejecting AGN model predictions. \textit{Bottom panel}: a UV diagnostic comparison of the rest-frame EW of \CIV\ with the \CIV/\CIII\ ratio \citep[from][]{2019MNRAS.487..333H}. This figure shows RCS0224z5 in comparison to several other extreme line emission galaxies (coloured data points, labelled in the bottom panel's legend): star-forming, high-redshift analogues in the local Universe \citep[; see text for details]{2017MNRAS.472.2608S, 2019ApJ...874...93B}, a LyC-emitting galaxy at $z \sim 3$ \citep{2020MNRAS.491.1093V}, two galaxies at $z \sim 6$ and $7$ shown to be star-forming \citep{2015MNRAS.454.1393S, 2017ApJ...836L..14M}, and a $z \sim 7$ galaxy likely dominated by an AGN \citep{2017ApJ...851...40L}. From these diagnostics, we conclude that there is no strong AGN activity present in RCS0224z5 and the emission is likely produced in star-forming \HII\ regions. Additionally, high-redshift galaxies exhibit \CIV\ EWs that are markedly higher than in any local analogues.}
	\label{fig:Line ratios CIV-HeII}
\end{figure}

\subsection{\texorpdfstring{\CIV}{CIV}: driven by star formation or AGN activity?}
\label{ssec:Discussion: CIV}

The high equivalent width emission of \CIV\ observed in RCS0224z5 ($\ssim 20 \, \Angstrom$) suggests the presence of a source producing ionising photons above the ionisation potential of \CIII, $47.9 \, \mathrm{eV}$ \citep[e.g.][]{2019ApJ...878L...3B}, as well as a high production efficiency of LyC photons, $\xi_\text{ion}$ \citep[e.g.][]{2015MNRAS.454.1393S}. In determining the origin of such ionising radiation -- star formation (SF) or AGN activity -- comparisons between observations and predictions from photoionisation models (e.g. \program{Cloudy}, see \citealt{2013RMxAA..49..137F}; \program{MAPPINGS}, see \citealt{2013ApJS..208...10D}) are now widely used \citep[e.g.][]{2001ApJ...556..121K, 2016MNRAS.462.1757G, 2016MNRAS.456.3354F, 2019A&ARv..27....3M}.

\Cref{fig:Line ratios CIV-HeII} examines the origin of the hard radiation field in RCS0224z5 with our constraints on the rest-frame UV line fluxes. The top panel of \cref{fig:Line ratios CIV-HeII} shows the \CIV/\CIII\ and \CIV/\HeII\ line ratios \citep[a SF vs. AGN diagnostic proposed by ][]{2016MNRAS.456.3354F}. We place lower limits on both ratios using the \CIV\ detection and upper limits on \CIII\ and \HeII. For comparison we show grids and lines for modelled nebular emission and narrow-line region AGN emission coloured according to their metallicity \citep[see legend; assuming a solar metallicity of $Z = 0.01524$,][]{2012MNRAS.427..127B}. The SF models are from \citet{2016MNRAS.462.1757G}, which are based on the latest version of the \citet{2003MNRAS.344.1000B} stellar population synthesis models, while the AGN models are drawn from \citet{2016MNRAS.456.3354F}. All models have a fixed hydrogen density ($n_\text{H}$; in this case $10^3 \, \mathrm{cm^{-3}}$ for AGN and $10^2 \, \mathrm{cm^{-3}}$ for SF models) and dust-to-heavy-element mass ratio (here, $\xi_\text{d}=0.3$). The SF models are shown for different values of ionisation parameter $U$, $\log_{10} U \in \left\{-4, -3, -2, -1 \right\}$, while the grids of narrow-line region AGN emission shown also have a varying $\alpha$ (ranging from $-2.0$ to $-1.2$), the power-law index of the specific luminosity $S_\nu \propto \nu^\alpha$ at rest-frame UV wavelengths, $\lambda_\text{emit} \leq 2500 \, \Angstrom$. We note that the results do not change significantly under different combinations of parameters $n_\text{H}$, $\xi_\text{d}$. The same is true for our assumption about the maximum electron density to establish the \CIII\ doublet flux (see \cref{ssec:Results: X-shooter}), as shown by the light grey measurement (reflecting a maximum of $n_e \leq 10^5 \, \mathrm{cm^{-3}}$ instead of $10^3 \, \mathrm{cm^{-3}}$; see \cref{ssec:Results: X-shooter}).

A confirmed LyC-emitting galaxy at $z \simeq 3.21$, Ion2, is shown for comparison \citep{2016A&A...585A..51D, 2016ApJ...825...41V, 2020MNRAS.491.1093V}; interestingly, its line ratios indicate a composite nature even though its spectral features in the UV have been attributed to young, massive stars \citep{2020MNRAS.491.1093V}. Also shown are two additional galaxies, at redshifts $6.11$ \citep{2015MNRAS.454.1393S} and $7.045$ \citep{2017ApJ...836L..14M} which are presumably similar to RCS0224z5, with low mass and a hard ionising radiation field found to be more likely originating from a metal-poor stellar population instead of an AGN. For these galaxies, and for RCS0224z5, the $2 \sigma$ upper limits strongly reject all possibilities of pure AGN models.

The photoionisation models by \citet{2016MNRAS.462.1757G} and \citet{2016MNRAS.456.3354F} are coupled to cosmological simulations by \citet{2019MNRAS.487..333H} to design \citetalias{1981PASP...93....5B}-like UV diagnostics to differentiate star-forming galaxies from AGN. These diagnostics are specifically designed to provide an accurate classification over a wide range of redshifts ($0 < z \lesssim 6$). Their diagnostic mapping the EW of \CIV\ against the \CIV/\HeII\ ratio is shown in the bottom panel of \cref{fig:Line ratios CIV-HeII}. We compare our measurement with local star-forming galaxies with extreme emission lines, that can therefore be seen as analogues of high-redshift galaxies: a sample selected through $\HeII \, \lambda \, 4687 \, \Angstrom$ emission \citep[implying a hard ionising spectrum; see][]{2017MNRAS.472.2608S}, and a sample of galaxies selected for high-EW $\OIIIf \, \lambda \, 5008 \, \Angstrom$ emission \citep[among other criteria;][]{2019ApJ...874...93B}. We show the high-redshift star-forming galaxies and mentioned above and, in contrast, a $z = 7.149$ galaxy showing evidence for AGN activity \citep{2017ApJ...851...40L}. The constraints on these high-redshift galaxies are in agreement with respectively the star formation and AGN models. However, there is a noticeable difference in their line ratios relative to the combined sample of low-redshift `analogues'.

From these diagnostics, we conclude that there is no strong AGN activity present in RCS0224z5 and the emission is likely produced in \HII\ regions. This finding agrees with \citet{2017MNRAS.467.3306S}, who concluded that the \CIV\ emission is likely nebular in origin, based on the `clumpy' \CIV\ morphology (as opposed to centrally concentrated emission). Instead, a young ($1$-$3 \, \mathrm{Myr}$), metal-poor stellar population likely has to account for the hard radiation field required for the \CIV\ emission, providing a considerable contribution of photons reaching energies of at least $47.9 \, \mathrm{eV}$. This fits into the picture of the prevalence of extreme line emitters in the early Universe -- both more common and more extreme than any known local analogues -- and accompanying hard radiation fields that has emerged recently \citep[e.g.][]{2014ApJ...784...58S, 2015ApJ...801..122S, 2015MNRAS.454.1393S, 2017ApJ...836L..14M, 2019ApJ...879...70H}.

\subsection{The \texorpdfstring{\NeIII/\OII}{[NeIII]/[OII]} line ratio as an ionisation and metallicity diagnostic alternative to \texorpdfstring{\OIIIf/\OII}{[OIII]/[OII]}}
\label{ssec:Discussion: NeIII/OII}

Neon is an $\upalpha$ element, produced by heavy stars ($M \gtrsim 8 \, \mathrm{M_\odot}$) in their carbon burning cycle and ultimately type II supernovae, and therefore tightly matches oxygen in abundance \citep[][, and references therein]{2019A&ARv..27....3M}. Moreover, \NeIII\ and its isoelectronic equivalent $\OIIIf \, \lambda \, 5008 \, \Angstrom$ (\OIIIf) have a similarly high ionisation potential ($41.0$ and $35.1 \, \mathrm{eV}$, respectively), meaning the luminosity ratio of \NeIII\ and the low-ionisation $\OII \, \lambda \, 3727, 3730 \, \Angstrom$ doublet (\OII) is a powerful diagnostic of the ionisation parameter, similar to \OIIIf/\OII\ \citep{2007MNRAS.381..125P, 2014ApJ...780..100L, 2019A&ARv..27....3M}. The observed relationship between metallicity and ionisation parameter also makes it a metallicity diagnostic, albeit indirectly \citep[causing it to exhibit a significant amount of scatter; see][]{2006A&A...459...85N, 2008A&A...488..463M}: metal-poor systems are expected to have a high \NeIII/\OII\ ratio.

The fact that \NeIII\ and \OII\ have both similar and short wavelengths gives the \NeIII/\OII\ diagnostic two distinct advantages over the widely used \OIIIf/\OII\ ratio: it is practically insensitive to dust attenuation, and it can be detected at higher redshifts with ground-based near-infrared instruments \citep{2014ApJ...780..100L}. The former has long been exploited \citep[e.g.][]{2002ApJ...581..205H}, and indeed, \NeIII\ has been detected several times at $z \gtrsim 3$, in combination with \OII: in particular, we will consider here the galaxies Ion2 at $z \simeq 3.21$, a confirmed LyC-emitting galaxy \citep[as discussed in \cref{ssec:Discussion: CIV};][]{2016A&A...585A..51D, 2016ApJ...825...41V, 2020MNRAS.491.1093V}, SMACS~J2031.8-4036 at $z \simeq 3.51$ \citep{2012MNRAS.427.1953C, 2012MNRAS.427.1973C, 2016MNRAS.456.4191P}, GOODSN-17940 at $z \simeq 4.41$ \citep{2017ApJ...846L..30S}, and LnA1689-2 at $z \simeq 4.87$ \citep{2014A&A...563A..58T}.

\begin{figure}
	\centering
	\includegraphics[width=\linewidth]{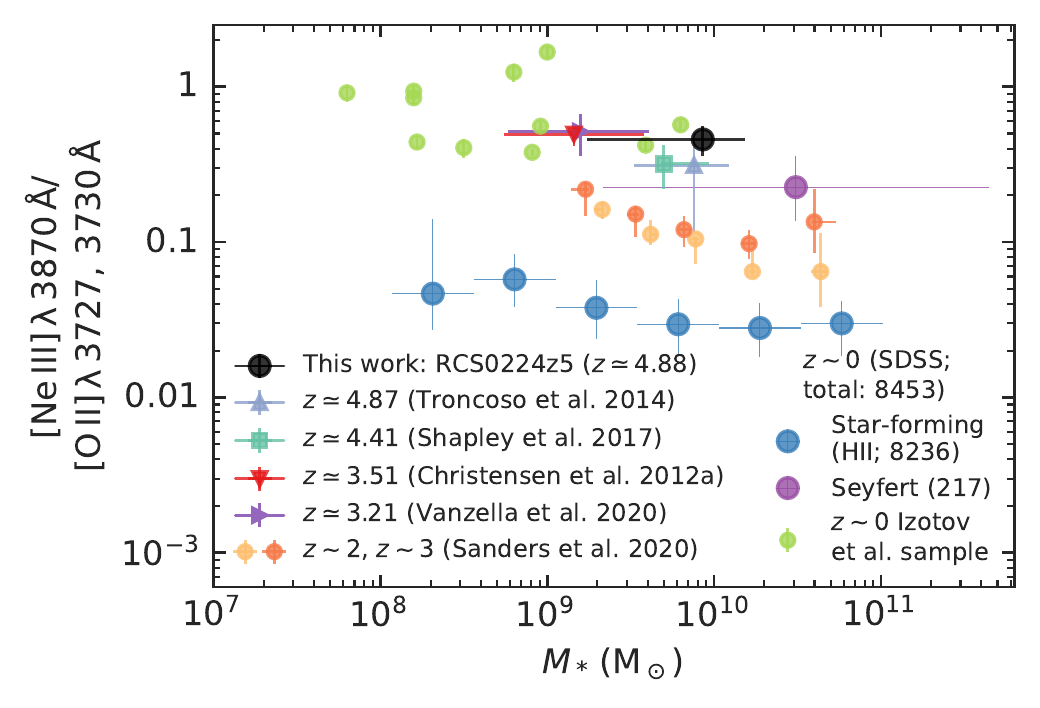}
	\caption[Measurements of \NeIII/\OII\ line ratios of local and high-redshift galaxies.]{\NeIII/\OII\ ratio of selected SDSS galaxies in bins of stellar mass (star-forming galaxies in blue, Seyfert in purple), as well as measurements of local LyC leakers (\Isample). Binned samples at $z = 2.3$ and $z = 3.3$ from the MOSDEF survey \citep{2021ApJ...914...19S} and the LyC-leaking galaxy Ion2 at $z \simeq 3.21$ \citep{2020MNRAS.491.1093V} are shown as a reference at intermediate redshift. At the very highest redshift for which measurements of \NeIII/\OII\ are currently possible, we include three star-forming galaxies, at $z \simeq 3.51$, $z \simeq 4.41$ and $z \simeq 4.87$ \citep[, respectively; see text]{2012MNRAS.427.1953C, 2017ApJ...846L..30S, 2014A&A...563A..58T}, and RCS0224z5. A tentative trend emerges, spanning nearly two orders of magnitude in the \NeIII/\OII\ ratio. Star-forming galaxies at $z \sim 0$ have a ratio of $\sim 0.01$ (with little scatter), whereas the ratio is significantly increased at higher redshift, $\ssim 0.1$ at $z \sim 2$-$3$. The trend culminates at $z \sim 5$: the ratio of RCS0224z5 ($\ssim 0.6$), seemingly typical based on this trend, is comparable to that of local LyC leakers, which are clear outliers compared to the general $z \sim 0$ galaxy population.
	}
	\label{fig:NeIII/OII line ratio vs stellar mass}
\end{figure}

\subsubsection{\texorpdfstring{\NeIII/\OII}{[NeIII]/[OII]} evolution over stellar mass and cosmic time}
\label{sssec:NeIII/OII evolution}

The emission line ratio $\NeIII/\OII = 0.46 \pm 0.10$ for RCS0224z5 is shown as a function of stellar mass in \cref{fig:NeIII/OII line ratio vs stellar mass}. To find the stellar mass for images 2 and 3 independent of magnification uncertainties, we follow the derivation of \citet{2007MNRAS.376..479S}, who reported a dynamical mass estimate of $M_\text{dyn} \sim 10^{10} \, \mathrm{M}_\odot$ based on the \OII\ velocity dispersion \citep[via Equation (1) in][]{2006ApJ...646..107E}. We assume a fiducial $C = 3 \pm 2$ \citep[with high uncertainty to reflect the range of possible values depending on mass distribution and velocity field, see][]{2006ApJ...646..107E}, and our measured \OII\ line width of $\sigma_\OII = 151 \pm 30 \, \mathrm{km/s}$ within $r = 2 \, \mathrm{kpc}$ \citep[following][]{2007MNRAS.376..479S}. We then take the typical stellar mass fraction of $z \sim 1$-$2$ star-forming galaxies of $27 \pm 5\%$, reflecting the range $22$-$32\%$ reported by \citet[; see also \citealt{2016ApJ...831..149W}]{2016MNRAS.457.1888S}, to derive a stellar mass of $(9 \pm 7) \times 10^{9} \, \mathrm{M}_\odot$. We note that with \textit{HST} photometry of image 1 (high magnification uncertainties prevent us from using the incomplete images 2 and 3), we derive a stellar mass of $4_{-3}^{+14} \times 10^8 \, \mathrm{M_\odot}$, using the \program{FAST} code \citep{2009ApJ...700..221K}, assuming a constant star formation rate (SFR), a \citet{2003PASP..115..763C} IMF, a minimum age of $10^7 \, \mathrm{yr}$ (the maximum age equal to the age of the universe at $z = 4.88$), an SMC dust law, the \citet{2003MNRAS.344.1000B} stellar libraries, and a metallicity of $Z = 0.2 \, \mathrm{Z_\odot}$. This estimate is marginally lower, but consistent within the uncertainty of the stellar mass obtained from the dynamical mass estimate.

Furthermore, the $\mathrm{SFR}_\OII$ of $12 \pm 2 \, \mathrm{M_\odot \, yr^{-1}}$ measured for image 2 and 3 by \citet{2007MNRAS.376..479S} -- combined with the stellar mass estimated from the dynamical mass -- places RCS0224z5 just under on the main sequence at its redshift \citep[e.g.][; a lower stellar mass, as suggested by the SED modelling, would shift it onto the main sequence]{2015ApJ...799..183S}. This supports the hypothesis that RCS0224z5 is a typical $z \sim 5$ star-forming galaxy.

To put our measurements of the near-UV \NeIII\ and \OII\ lines into perspective, we turn to a large observational sample of nearby galaxies from the Sloan Digital Sky Survey Data Release 7 \citep[SDSS DR7;][]{2009ApJS..182..543A}, retrieving line fluxes from the MPA-JHU emission line catalogue for $\num[group-separator={,}]{827640}$ unique sources.\footnote{The catalogue is available at \url{https://www.mpa-garching.mpg.de/SDSS/DR7/}, while a description of the relevant methods used to compile it can be found in \citet{2004ApJ...613..898T}.} A detailed description of the selection procedure for the galaxies used here is given in \cref{ap:SDSS selection}. The \NeIII/\OII\ ratio of these nearby galaxies are shown in bins of stellar mass (for bins with at least $25$ galaxies), for the two main classes (star-forming and Seyfert; see \cref{ap:SDSS selection}). Additionally, we compare these to local LyC-leaking galaxies; in particular, we will consider a sample compiled from \citet{2016MNRAS.461.3683I, 2016Natur.529..178I, 2018MNRAS.474.4514I, 2018MNRAS.478.4851I, 2020A&A...639A..85G, 2020MNRAS.497.4293G}, together simply the \Isample\ hereafter. Finally, individual measurements of the aforementioned high-redshift galaxies at $z \simeq 3.21, 3.51, 4.41, 4.87$ \citep[, respectively]{2020MNRAS.491.1093V, 2012MNRAS.427.1953C, 2017ApJ...846L..30S, 2014A&A...563A..58T} are shown.

\Cref{fig:NeIII/OII line ratio vs stellar mass} illustrates that RCS0224z5 has a \NeIII/\OII\ ratio consistent with LyC leakers (both those at low redshift and Ion2 at $z \simeq 3.21$), and nearly two orders of magnitude higher than local star forming galaxies with the same mass. This agrees with recent findings of enhanced \NeIII/\OII\ ratios in star-forming galaxies at $z \sim 2$ \citep{2015ApJ...798...29Z, 2020ApJ...902L..16J}. Moreover, this is in agreement with the expectation that \NeIII/\OII\ is a proxy of \OIIIf/\OII, and that RCS0224z5 might have a high LyC escape fraction, as discussed further below (\cref{ssec:Discussion: Lya and MgII}).

\begin{figure}
	\centering
	\includegraphics[width=\linewidth]{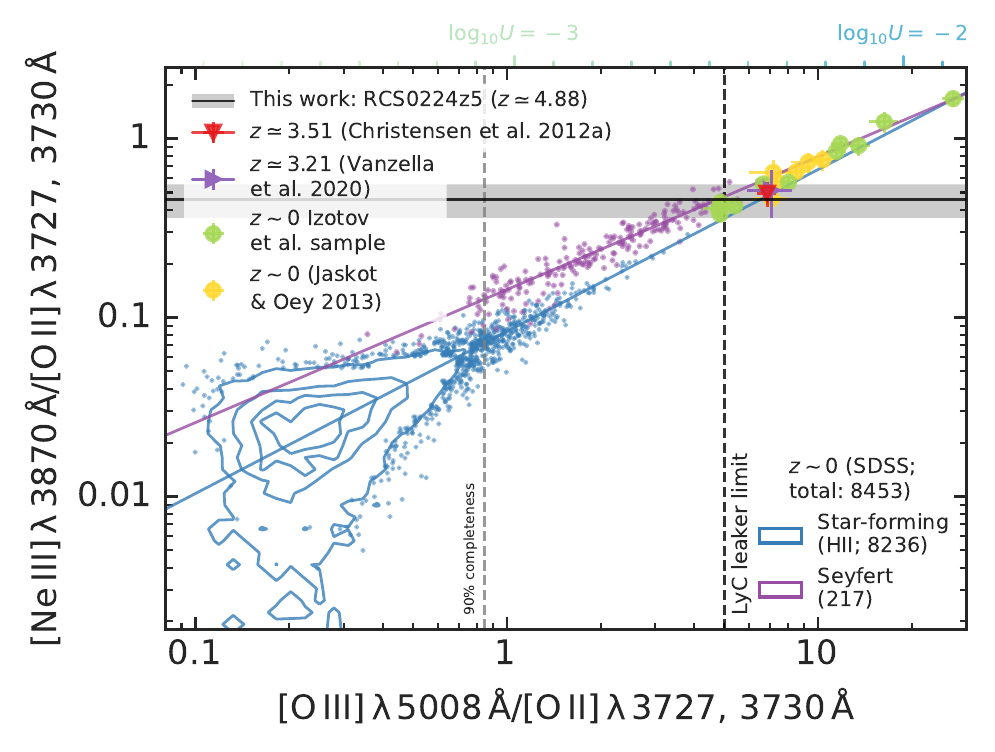}
	\caption[Correlation between the \NeIII/\OII\ and \OIIIf/\OII\ line ratios of SDSS galaxies.]{Correlation between the \NeIII/\OII\ and \OIIIf/\OII\ line ratios of SDSS galaxies (star-forming galaxies in blue, Seyfert in purple). Also shown are Ion2 at $z \simeq 3.21$ \citep{2020MNRAS.491.1093V}, SMACS~J2031.8-4036 at $z \simeq 3.51$ \citep{2012MNRAS.427.1953C}, the LyC leakers from the \Isample, as well as $6$ extreme ``Green Pea'' galaxies from \citet{2013ApJ...766...91J}. The vertical grey dashed line shows the completeness limit of 90\%, while the fiducial limit ratio of $\OIIIf/\OII \geq 5$ used to characterise LyC leakers is shown by the vertical black dashed line (see text for details on both). Ionisation parameter values of $\log_{10} U = -3, -2$ are indicated by vertical blue lines \citep[see \cref{eq:OIII/OII logU diagnostic}, derived by][]{2000MNRAS.318..462D}. Direct measurements of Ion2 and SMACS~J2031.8-4036 suggest the correlation between \NeIII/\OII\ and \OIIIf/\OII\ seen in local galaxies holds well at higher redshift. Both these sources and RCS0224z5, shown by the black line with grey uncertainty, have a high ionisation parameter compared to local galaxies, similar to that of LyC-leaking systems, but may very well be typical examples at high redshift as they do not seem to be as extreme outliers compared to their contemporaries (\cref{fig:NeIII/OII line ratio vs stellar mass}).}
	\label{fig:NeIII/OII vs OIII/OII line ratios}
\end{figure}

\subsubsection{\texorpdfstring{\NeIII/\OII}{[NeIII]/[OII]} as a proxy for the ionisation state of the ISM}
\label{sssec:NeIII/OII as a proxy for the ionisation state of the ISM}

As evidenced by its tight correlation with \OIIIf/\OII\ in local galaxies \citep[e.g.][]{2020ApJ...902L..16J}, the \NeIII/\OII\ ratio is a robust tracer of the ionisation parameter. Previous studies have proved the strong similarity between the \NeIII/\OII\ and \OIIIf/\OII\ ratios theoretically; however, their quantitative predictions did not quite match observations, which is thought to be due to the modelled ionising spectra being insufficiently hard \citep{2014ApJ...780..100L}. We therefore construct an empirical relationship here by exploiting the statistics of SDSS. In \cref{fig:NeIII/OII vs OIII/OII line ratios}, the distributions of \NeIII/\OII\ and \OIIIf/\OII\ line ratios are shown for two classes of SDSS galaxies (star-forming and Seyfert; see \cref{ap:SDSS selection}). The LyC leakers from the \Isample\ are again shown for comparison, along with six other local galaxies selected for extreme \OIIIf\ emission \citep[``Green Peas'', or GPs;][]{2013ApJ...766...91J}, as well as the high-redshift galaxies Ion2 and SMACS~J2031.8-4036 \citep[, respectively]{2020MNRAS.491.1093V, 2012MNRAS.427.1953C}. The fiducial limiting ratio of $\OIIIf/\OII \geq 5$ used to characterise LyC leakers is shown by a vertical dashed black line.

To fit the relationship from the SDSS sample, we first define the completeness of a given \OIIIf/\OII\ bin to be the fraction of galaxies contained in that bin with a \NeIII\ SNR higher than 5. At values of $\OIIIf/\OII \lesssim 1$, this completeness drops significantly. Since we require a highly significant \OII\ detection ($\text{SNR} > 30$, see \cref{ap:SDSS selection}), the uncertainty on the \NeIII/\OII\ ratio is dominated by that of \NeIII, explaining the scattered cloud of points in contrast to the tight correlation seen at $\OIIIf/\OII > 1$. As \NeIII\ is the weakest out of the three lines considered here \citep{2019A&ARv..27....3M}, we fit the relation taking into account uncertainties on the \NeIII/\OII\ ratio only. For star-forming galaxies we exclude points in the region where the completeness fails to meet $90\%$ ($\OIIIf/\OII < 0.849$; grey dashed line in \cref{fig:NeIII/OII vs OIII/OII line ratios}). For the remaining $272$ points, we find a Spearman's rank correlation coefficient (measured in log-log space) of $\rho_\text{S} = 0.87$, indicating a strong positive correlation. Moreover, the resulting fit captures the behaviour at both the low and high \OIIIf/\OII\ ratios well: extrapolating to lower \OIIIf/\OII\ ratios the low-SNR SDSS data are scattered around the relation symmetrically, while extrapolating to higher \OIIIf/\OII\ ratios the extreme ratios of GP galaxies (not included in the fit) are consistent with this trend. The fit for star-forming galaxies is given by
\begin{equation}
    \label{eq:NeIII-OII vs OIII/OII correlation}
    \log_{10} \left( \frac{\NeIII}{\OII} \right) = 0.9051 \log_{10} \left( \frac{\OIIIf}{\OII} \right) - 1.078
\end{equation}

For example, the proposed line ratio above which a significant fraction of sources might be leaking LyC photons, $\OIIIf/\OII \geq 5$, \citep[e.g.][]{2016Natur.529..178I, 2016MNRAS.461.3683I} corresponds to $\NeIII/\OII \geq 0.36$. Furthermore, with this empirical relationship, we can translate diagnostics based on the \OIIIf/\OII\ ratio, e.g. a diagnostic for the ionisation parameter \citep[derived from single-star photoionisation models, see][]{2000MNRAS.318..462D},
\begin{equation}
    \label{eq:OIII/OII logU diagnostic}
    \log_{10} U = 0.80 \log_{10} \left( \frac{\OIIIf}{\OII} \right) - 3.02
\end{equation}

\noindent Vertical blue lines indicate values of $\log_{10} U = -3, -2$ in \cref{fig:NeIII/OII vs OIII/OII line ratios}. Combining \cref{eq:NeIII-OII vs OIII/OII correlation,eq:OIII/OII logU diagnostic}, we derive
\begin{align}
    \label{eq:NeIII/OII logU diagnostic}
    \log_{10} U & = 0.80 \frac{\log_{10} \left( \NeIII/\OII \right) + 1.078}{0.9051} - 3.02 \nonumber
    \\
    & = 0.884 \log_{10} \left( \frac{\NeIII}{\OII} \right) - 2.07
\end{align}

\noindent In the case of RCS0224z5, we find an ionisation parameter of $\log_{10} U = -2.37 \pm 0.08$. Given its derivation from the \NeIII/\OII\ ratio, we again note this is likely not an extreme case at its redshift \citep[\cref{fig:NeIII/OII line ratio vs stellar mass}, and see similar estimates of $U$ at $z \sim 7$-$8$ in][]{2017MNRAS.464..469S}, but a high ionisation parameter compared to local galaxies, similar to that of LyC-leaking systems (the limit of $\OIIIf/\OII \geq 5$ translates to $\log_{10} U = -2.46$).

However, the \NeIII/\OII\ ratio is also an indirect tracer of the gas metallicity (being mostly anti-correlated with metallicity). The fact that RCS0224z5 (along with GOODSN-17940) has \NeIII/\OII\ much higher than local galaxies, and hence much lower metallicity, is also likely linked to the redshift evolution of the mass-metallicity relation (MZR). This is illustrated in \cref{fig:NeIII/OII line ratio vs stellar mass} where binned samples are shown at intermediate redshifts, $z = 2.3$ and $z = 3.3$ \citep[measurements from the MOSDEF survey;][]{2021ApJ...914...19S}. The tentative evolutionary trend of the \NeIII/\OII\ ratio with redshift seen in \cref{fig:NeIII/OII line ratio vs stellar mass} has indeed been verified to approximately reproduce the MZR as observed with a variety of metallicity tracers \citep[e.g.][]{2008A&A...488..463M}. The next section will discuss the diagnosticity of \NeIII/\OII\ specifically regarding metallicity in further detail.

\begin{figure}
	\centering
	\includegraphics[width=\linewidth]{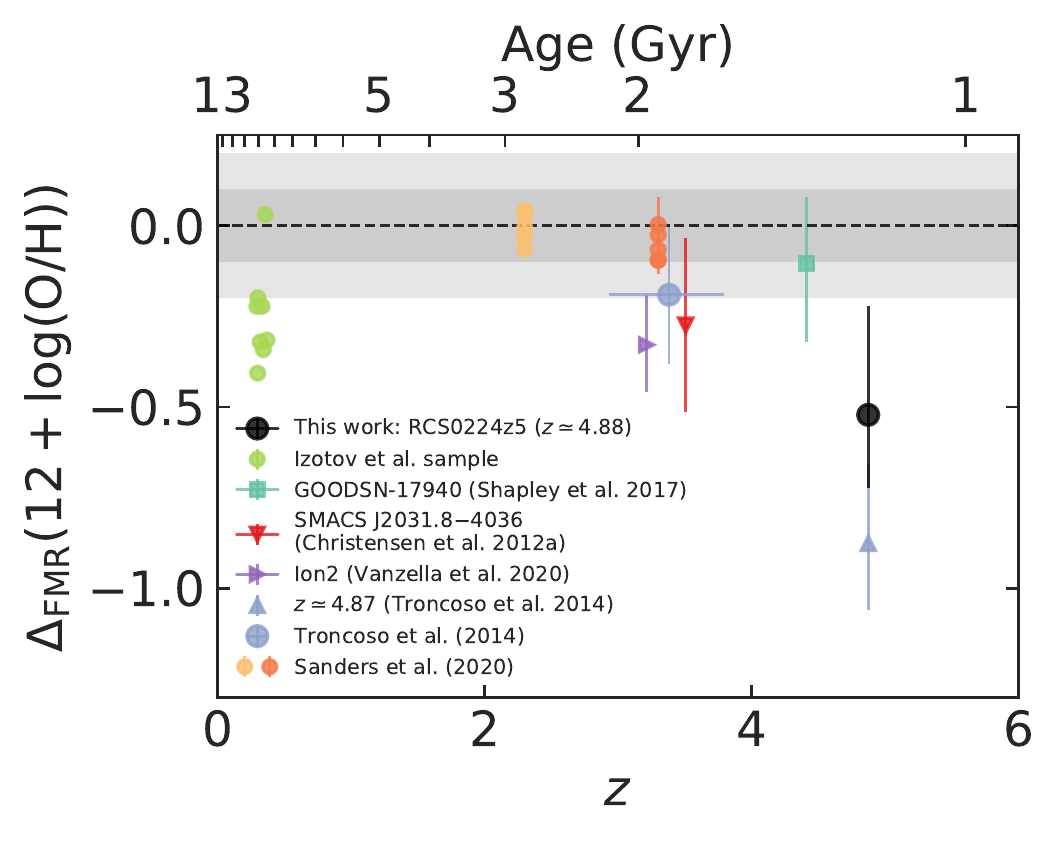}
	\caption[Offsets from the FMR.]{Offsets from the Fundamental Metallicity Relation (FMR), given by the inferred metallicity minus the one derived from the FMR. Intrinsic uncertainty of the FMR (which varies with both stellar mass and SFR) is indicated by the grey shaded regions, the darker region typical for galaxies at high mass and low SFR, the lighter region for galaxies at low mass and high SFR \citep[for details, see][]{2020MNRAS.491..944C}. At low redshift, the LyC leakers from the \Isample\ are shown. Samples binned by stellar mass at $z = 2.3$ and $z = 3.3$ from the MOSDEF survey \citep{2021ApJ...914...19S} are shown, as well as a binned sample of $31$ galaxies at $z \sim 3$-$4$ from the AMAZE and LSD surveys, presented in \citet{2014A&A...563A..58T}. LnA1689-2 is excluded from the latter and instead shown individually at $z \simeq 4.87$, as are Ion2 at $z \simeq 3.21$ \citep{2020MNRAS.491.1093V}, SMACS~J2031.8-4036 at $z \simeq 3.51$ \citep{2012MNRAS.427.1953C}, GOODSN-17940 at $z \simeq 4.41$ \citep{2017ApJ...846L..30S}. Galaxies at the highest redshift seem to show a trend towards negative differences, i.e. metallicities smaller than expected from the FMR.}
	\label{fig:NeIII/OII delta FMR}
\end{figure}

\subsubsection{Tracing metallicity with \texorpdfstring{\NeIII/\OII}{[NeIII]/[OII]}}
\label{sssec:Tracing metallicity with NeIII/OII}

By virtue of its correlation with the ionisation parameter, the \NeIII/\OII\ ratio is also an indirect tracer of the metallicity. In the following we discuss the former aspect in greater detail, albeit with the caveat that the ratio is a more robust diagnostic for the ionisation parameter than for metallicity.

By using the metallicity diagnostic relation for \NeIII/\OII\ from \citet{2018ApJ...859..175B}, calibrated with local analogues of high-redshift galaxies, we obtain $12 + \log \left ( \text{O/H} \right) = 8.01_{-0.21}^{+0.21}$ for RCS0224z5, where we have included a $0.2 \, \mathrm{dex}$ systematic calibration uncertainty (see e.g. \citealt{2006A&A...459...85N}). This corresponds to roughly $20\%$ of the solar metallicity: $Z = 0.21_{-0.08}^{+0.13} \, \mathrm{Z_\odot}$.\footnote{We adopt a solar oxygen abundance of $12 + \log \left ( \text{O/H} \right)_\odot = 8.69$ \citep{2009ARA&A..47..481A}.} The galaxies at $z \sim 4$-$5$ with \NeIII\ and \OII\ measurements are clearly characterised by significantly higher line ratios (\cref{fig:NeIII/OII line ratio vs stellar mass}), and hence higher ionisation parameters than $z \sim 3.5$ galaxies. Using the (uncertain) strong-line calibration, we can indirectly infer they have lower metallicities too, which would confirm that they do follow the general redshift evolution of metallicity.

However, one should take into account that the metallicity of star-forming galaxies also shows a secondary dependence on SFR that, together with the primary dependence on mass, is dubbed Fundamental Metallicity Relation (FMR, see e.g. \citealt{2010MNRAS.408.2115M}; see also a review in \citealt{2019A&ARv..27....3M}). This secondary dependence is thought to result from a more fundamental relation with the gas content \citep{2013MNRAS.433.1425B, 2016MNRAS.455.1156B, 2016A&A...595A..48B} and ascribed primarily to the accretion of near-pristine (or low-metallicity) gas which increases the gas content and dilutes the metallicity; the increased gas content also results into an increased SFR through the Kennicutt–Schmidt relation, which gives the observed anti-correlation between SFR and metallicity. Once this secondary dependence is taken into account then the redshift evolution of the metallicity is essentially absent, at least out to $z \sim 2.5$ \citep{2010MNRAS.408.2115M, 2019A&A...627A..42C}. Some deviation from the FMR was claimed at $z > 3$ \citep{2014A&A...563A..58T}, but more recently \cite{2021ApJ...914...19S} have shown that their sample at z$\sim$3 follows the same FMR as local galaxies if metallicities are measured through the \citet{2018ApJ...859..175B} calibration.

We assess whether RCS0224z5 and other high-redshift galaxies, as well as the local LyC leakers from the \Isample, follow the FMR, carefully revisiting all measurements in a consistent way, as discussed in the following. At high redshift, we consider galaxies at $z > 2$ from \citet{2021ApJ...914...19S} and \citet{2014A&A...563A..58T}, as well as the three galaxies at $z > 4$ with detections of \NeIII\ and \OII\ already discussed in the previous section. For RCS0224z5 we use the stellar mass of images 2 and 3 obtained from the dynamical mass, $(9 \pm 7) \times 10^{9} \, \mathrm{M}_\odot$ (including uncertainties on $C$, the stellar mass fraction, and $\sigma_\OII$, discussed in \cref{sssec:NeIII/OII evolution}), and the SFR of images 2 and 3, $12 \pm 2 \, \mathrm{M_\odot \, yr^{-1}}$, as reported by \citet{2007MNRAS.376..479S}. We also include the two individual galaxies at $z \sim 3$ with \NeIII\ and \OII\ measurements discussed in the previous section. We use the same high-redshift calibration from \citet{2018ApJ...859..175B} for all high-redshift galaxies. We then consider $\Delta_\text{FMR} ( 12 + \log ( \text{O/H} ))$, the deviation of the measured metallicity from the local FMR, defined as
\begin{align*}
    \Delta_\text{FMR} \left( 12 + \log \left ( \text{O/H} \right) \right) = & 12 + \log \left ( \text{O/H} \right)
    \\
    & - \left( 12 + \log \left ( \text{O/H} \right) \right)_\text{FMR}.
\end{align*}

\noindent Here, we describe the metallicity predicted by the local FMR, $( 12 + \log ( \text{O/H} ))_\text{FMR}$, as a function of stellar mass and SFR through Equation (5) in \citet{2020MNRAS.491..944C}. The best-fit parameters obtained by \citeauthor{2020MNRAS.491..944C} adopt $T_e$-based metallicity calibrations \citep[as][, but based on the full SDSS dataset]{2018ApJ...859..175B}. Uncertainties are estimated by independently varying the input variables $(M_*, \text{SFR})$ within the $1 \sigma$ uncertainty range and adding the resulting deviations in quadrature. For LyC leakers in the \Isample, we calculate the FMR offset if the metallicity has been reported. We show the resulting deviations from the FMR in \cref{fig:NeIII/OII delta FMR}.

While at $z \sim 2$ galaxies are fully consistent with the local FMR, at $z > 3$ galaxies start having a larger scatter and tend to be distributed towards lower metallicities with respect to the FMR, also depending on the sample. Although the uncertainties are large, RCS0224z5 also shows some mild tension with respect to the local FMR, being more metal poor. Interestingly, the nearby LyC-leaking galaxies exhibit a significant offset in a similar direction. The other galaxy at nearly the same redshift, LnA1689-2 from the \citeauthor{2014A&A...563A..58T} sample, likewise deviates from the FMR. Since the FMR is considered a relation describing the smooth evolution of galaxies in near-equilibrium between the inflow and outflow of gas and star formation, these findings may suggest that such young galaxies at $z \sim 5$ are in an early stage of evolution in which they have not yet reached a steady equilibrium as the more evolved galaxies at lower redshifts, possibly as a consequence of an excess in gas accretion, which results in additional dilution of metals. However, these results should be confirmed with a larger sample of galaxies at $z > 4$ and by using additional metallicity diagnostics, which will certainly be feasible with \textit{JWST}.

Finally, we note that the estimated metallicity obtained through the \NeIII/\OII\ ratio is significantly higher than the values required in stellar population synthesis models to reproduce the observed \CIV\ EW ($Z < 0.05 \, \mathrm{Z_\odot}$, as shown in \citealt{2017MNRAS.467.3306S}). This indicates the hardness of the ionising spectrum may currently be underestimated in such models. The lack of ionising photons above $47.9 \, \mathrm{eV}$ could be accounted for by physics currently not captured in models \citep[e.g. stars stripped in binaries,][]{2019A&A...629A.134G}, or it may be explained by a lower stellar iron abundance than derived using the nebular oxygen abundance and assuming solar oxygen-to-iron ratios (O/Fe). A higher oxygen-to-iron is expected in the early Universe compared to local galaxies, given that AGB stars have lifetimes of only a few gigayears and metal enrichment is dominated by supernovae \citep{2019A&ARv..27....3M}. In particular, the discrepancy seems to agree well with the findings of \citet{2016ApJ...826..159S}, who demonstrate that the oxygen-to-iron ratio of their sample of star-forming galaxies at $z \sim 2.4$ is elevated by a factor of $\ssim 4$ relative to the solar value (i.e. virtually the same enhancement as the ratio between the metallicities discussed here, $20\%$ and $5\%$ solar). While RCS0224z5 appears moderately metal-enriched as measured indirectly with \NeIII/\OII\ (probing the nebular oxygen abundance), stellar atmospheres could be significantly more iron-poor than expected when assuming a solar O/Fe ratio, resulting in an ionising spectrum sufficiently hard to explain the observed \CIV\ EW.

More generally, these results show the potential of \NeIII\ as a powerful diagnostic, specifically for the study of high-redshift galaxies. Note, however, that the \NeIII\ line can blend with $\ion{He}{I} \, \lambda \, 3890 \, \Angstrom$, separated by $\ssim 1500 \, \mathrm{km/s}$, in low-resolution spectra ($R \lesssim 200$). This effect becomes more prominent when the lines are broadened, for example with a significant contribution from the broad-line region of an AGN \citep[e.g.][]{2017ApJ...846..102M} -- this will not be the case, however, for star-forming systems dominated by narrow nebular emission lines.

\begin{figure}
	\centering
	\includegraphics[width=\columnwidth]{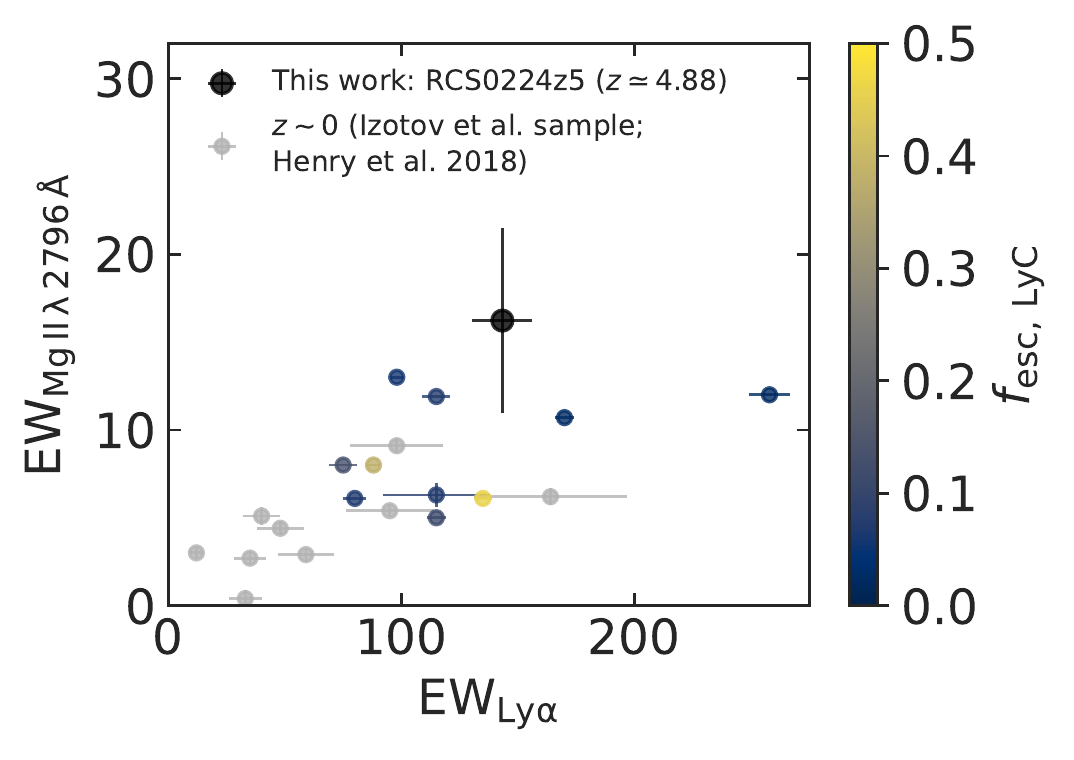}
	\caption[EWs of \lya\ and $\MgII \, \lambda \, 2796 \, \Angstrom$.]{Equivalent widths of \lya\ and $\MgII \, \lambda \, 2796 \, \Angstrom$ of RCS0224z5, compared to local extreme emission line galaxies, among which ten LyC-emitting sources \citep[the \Isample\ and][, coloured according to the escape fraction of LyC, if known]{2018ApJ...855...96H}. The correlation between \lya\ and \MgII\ EWs is one of the indicators that the escape mechanisms of \lya\ and \MgII\ are similar. Moreover, relatively high EWs imply that both \lya\ and \MgII\ escape is comparatively high in RCS0224z5.
	}
	\label{fig:Lya-MgII EW distribution}
\end{figure}

\begin{figure}
	\centering
	\includegraphics[width=\columnwidth]{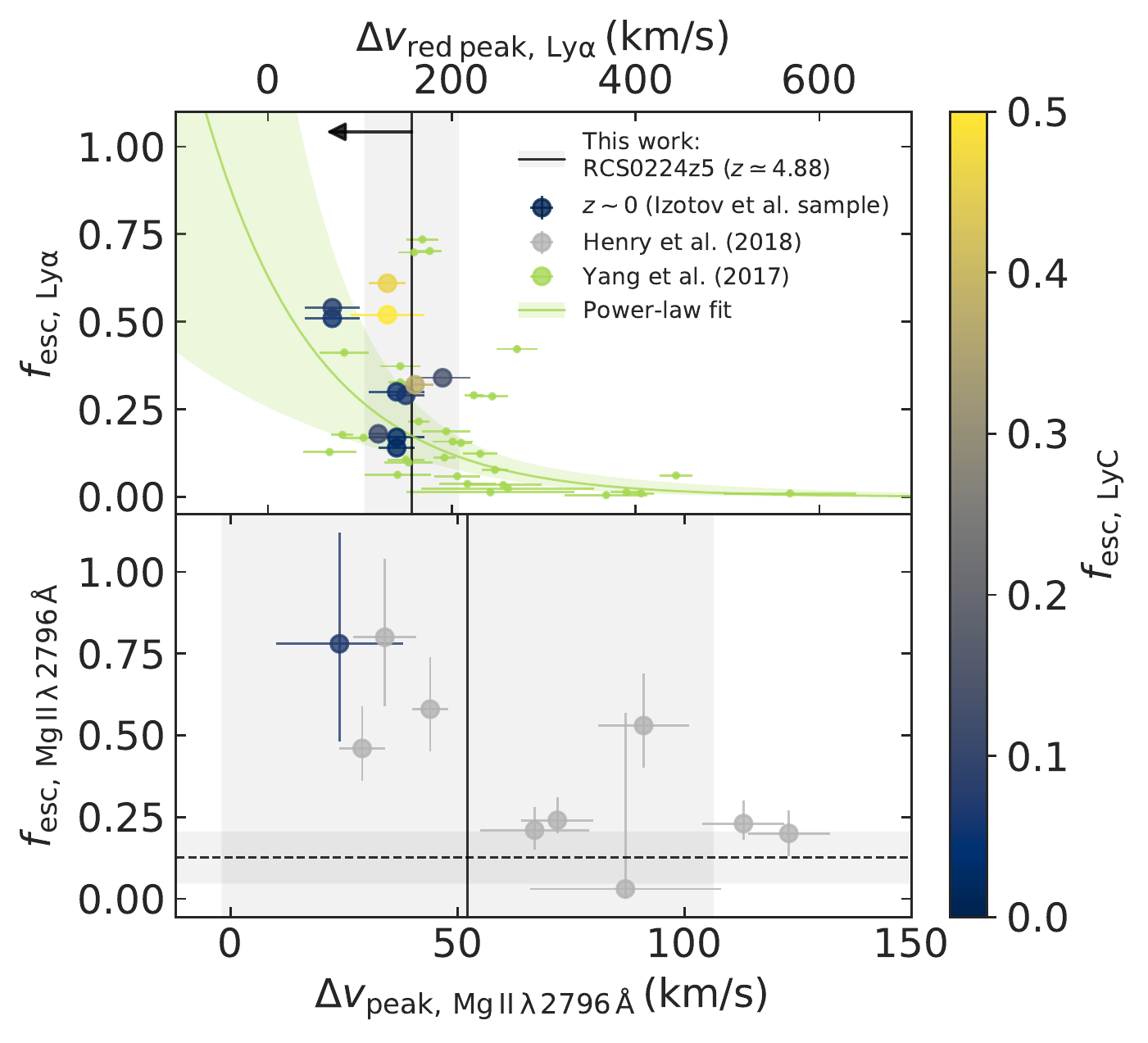}
	\caption[Escape fractions compared to velocity offsets of \lya\ and $\MgII \, \lambda \, 2796 \, \Angstrom$.]{Escape fractions of \lya\ radiation (top) and $\MgII \, \lambda \, 2796 \, \Angstrom$ (bottom) compared to the corresponding velocity offsets of \lya\ (its red peak; top) and the peak of $\MgII \, \lambda \, 2796 \, \Angstrom$ (bottom), compared to local extreme emission line galaxies. Among these are ten LyC-emitting sources \citep[the \Isample\ and][, coloured according to the escape fraction of LyC, if known]{2018ApJ...855...96H}. In the top panel, a sample of Green Pea galaxies from \citet{2016ApJ...820..130Y, 2017ApJ...844..171Y} is shown (the sample consists of $43$ galaxies, among which $5$ are already contained in the \Isample), with a simple power-law fit. Both panels show a similar trend (though with considerable scatter), where a small velocity offset indicates a large escape fraction. The measured velocity offsets of RCS0224z5 (effectively an upper limit in the case of {\lya} due to IGM absorption) again suggest relatively high escape fractions, which is supported by the predicted escape fraction of $\MgII \, \lambda \, 2796 \, \Angstrom$ (the horizontal dashed line; see text for details).
	}
	\label{fig:Lya/LyC-MgII escape and velocity offsets}
\end{figure}

\subsection{LyC escape traced by \texorpdfstring{\lya\ and \MgII}{\lyatext and MgII} emission}
\label{ssec:Discussion: Lya and MgII}

\subsubsection{Velocity offsets of \texorpdfstring{\lya}{\lyatext} and {\MgII} }
\label{sssec:Lya}

Using spatially resolved MUSE spectroscopy, \citet{2017MNRAS.467.3306S} demonstrated the presence of extended, high-EW \lya\ emission, with a narrow, red peak emerging very close to the systemic redshift (defined by \OII), regardless of position within the \lya\ halo. X-shooter has a higher spectral resolution than MUSE at $\ssim 7000 \, \Angstrom$ ($R \sim 6500$ versus $R \sim 2700$, respectively). We fit an asymmetric Gaussian line profile \citep[see e.g.][]{2014ApJ...788...74S} to the \lya\ observed in image 1 by X-shooter (see \cref{fig:Overview panel}). We find a velocity offset of $\Delta v_\text{peak} = 156 \pm 52 \, \mathrm{km/s}$ from the systemic redshift \citep[in agreement with][]{2017MNRAS.467.3306S}, FWHM of $207 \pm 6 \, \mathrm{km/s}$, and asymmetry factor $a_\text{asym} = 0.32 \pm 0.01$, indicating a skewed profile with a red wing. The asymmetric Gaussian provides an overall good fit, but does show non-vanishing residuals, particularly at the line peak, as well as the regions around $400$ and $1000 \, \mathrm{km/s}$. As we lack sufficient angular resolution in the spectrum to study the \lya\ line shape in depth, we will not focus on possible interpretations here and instead refer to the extensive discussion in \citet[][]{2017MNRAS.467.3306S}.

The determination of the peak offset can provide interesting constraints on properties of the neutral ISM. In particular, the peak \emph{separation} in a double-peaked \lya\ profile (peak offset being the closest alternative to peak separation in the absence of a blue peak\footnote{The velocity offset of the blue peak actually offers the best predictive power \citep{2017A&A...597A..13V}.} in the observed spectrum as a result of absorption in the intervening neutral IGM) is an indicator of resonant scattering of escaping \lya\ photons: a higher column density of neutral hydrogen along the path of escape means more scattering, affecting the line profile shape to have its peak appear increasingly further from the systemic velocity at which the photons were produced: only photons far away from resonance (in frequency space), with a resulting low cross section, are able to escape \citep[e.g.][]{2017A&A...597A..13V}. In the context of the escape of LyC, which is itself governed by the same distribution of neutral hydrogen, the peak separation has proven to be a solid predictor of LyC escape fractions \citep{2017A&A...597A..13V, 2018MNRAS.478.4851I, 2020A&A...639A..85G}. There are other indirect probes of LyC escape, though, which become necessary for application in the EoR, where \lya\ (including the red peak) can be fully absorbed, as a result of the broad damping wing \citep[e.g.][]{2014PASA...31...40D}. Within this context, \MgII, also a resonant transition in the near UV, presents promising features for high-redshift studies.

We detect $\MgII \, \lambda \, 2796 \, \Angstrom$ as a narrow emission line close to the systemic redshift ($\ssim 52 \, \mathrm{km/s}$, see \cref{fig:Overview panel} and \cref{tab:Results}), while this feature is commonly seen in absorption in the spectra of local galaxies. \MgII\ P-Cygni profiles with blueshifted absorption and redshifted emission have been discovered in more distant ($z \gtrsim 0.5$) galaxies, however, where it has been exploited to study galactic outflows \citep[e.g.][]{2009ApJ...692..187W, 2010ApJ...719.1503R, 2011ApJ...728...55R, 2011ApJ...743...95G, 2012ApJ...759...26E, 2013ApJ...774...50K, 2016MNRAS.458.1891B, 2017A&A...608A...7F}. Previous studies of gravitationally lensed galaxies at $z \sim 1.5$-$2$ have also demonstrated several cases of \MgII\ emission, seen without P-Cygni profiles or evidence for a redshift from the systemic velocity \citep{2010MNRAS.402.2335P, 2014ApJ...790...44R, 2016A&A...585A..27K}.

Nebular emission from \HII\ regions or resonant scattering (of either \MgII\ or continuum photons) are both plausible sources of \MgII\ emission. The former scenario is supported by comparing observed \MgII\ profiles to photoionisation models, while their variety, ranging from narrow, systemic emission to P-Cygni and pure absorption, provides evidence for the latter \citep{2010ApJ...719.1503R, 2012ApJ...759...26E, 2018A&A...617A..62F}. For RCS0224z5, its narrow line profile observed close to the systemic redshift suggests a nebular origin of the \MgII\ line and an ISM where little scattering takes place along the line of sight (e.g. due to a very high filling factor of ionised gas).

\subsubsection{The predictive power of \texorpdfstring{\lya\ and \MgII}{\lyatext and MgII} emission for unseen LyC escape processes}
\label{sssec:Lya and MgII tracing LyC escape}

Considering the resonant nature of \MgII\ (and the low ionisation energies of magnesium that make \MgII\ mainly a tracer of the neutral ISM), various studies have pointed out the resemblance to \lya\ \citep[e.g.][]{2018ApJ...855...96H, 2018A&A...617A..62F}. As a result, \MgII\ might provide a new way to indirectly but effectively identify sources emitting both \lya\ \citep[commonly used as a probe for measuring the conditions of the IGM, see][]{2014PASA...31...40D} and LyC radiation at $z \gtrsim 6$, before reionisation is completed \citep{2015MNRAS.447..499M}, where neither may be directly observable as a result of absorption by the neutral IGM. Indeed, \MgII\ emission has been reported in ten local LyC leakers \citep{2016MNRAS.461.3683I, 2016Natur.529..178I, 2018MNRAS.474.4514I, 2018MNRAS.478.4851I, 2020MNRAS.497.4293G}, and in a sample of Green Pea (GP, see also \cref{sssec:NeIII/OII as a proxy for the ionisation state of the ISM}) galaxies \citep{2018ApJ...855...96H}. In the latter, a correlation between the escape fractions of \MgII\ and \lya\ has been found, which suggests a tentative correlation between the escape fractions of \MgII\ and LyC. If well established \citep[promising first results have been reported,][]{2020MNRAS.498.2554C, 2021MNRAS.505.1382M}, this correlation could allow \textit{JWST} and the generation of Extremely Large Telescopes to reveal the sources beyond (and behind) reionisation, given they can detect a \MgII\ signal from sources in the EoR. Moreover, \MgII\ would provide a tool to predict the intrinsic properties of \lya\ within galaxies, in order to improve the constraints on the neutral fraction in the IGM derived from the \lya\ prevalence in reionisation-era sources \citep[e.g.][]{2014ApJ...795...20S, 2018ApJ...856....2M, 2018A&A...619A.147P}.

Using the measured EW of $\MgII \, \lambda \, 2796 \, \Angstrom$ we compare RCS0224z5 to local LyC leakers and GPs from the \citeauthor{2016Natur.529..178I} and \citeauthor{2018ApJ...855...96H} samples in \cref{fig:Lya-MgII EW distribution}. This figure illustrates the correlation between \lya\ and \MgII\ EWs (although lacking an evident direct relation with LyC escape fraction shown by the colourbar), as would be expected for correlated escape fractions of the two lines.

As discussed in \cref{sssec:Lya}, the velocity offsets of a resonant emission line can be a proxy for the fraction of escaping photons (i.e. both \lya\ and \MgII). \Cref{fig:Lya/LyC-MgII escape and velocity offsets} therefore compares escape fractions with velocity offsets, respectively \lya\ escape as a function of the velocity offset of its red peak, and $\MgII \, \lambda \, 2796 \, \Angstrom$ escape as a function of peak velocity offset of that same line. In the case of \lya, an additional sample of Green Pea galaxies from \citet{2016ApJ...820..130Y, 2017ApJ...844..171Y} is shown for reference (the sample consists of $43$ galaxies, among which $5$ are already contained in the \Isample). The data points show a significant amount of scatter, but the velocity offsets are negatively correlated with the escape fractions. We fit a simple power-law of the form
\begin{equation}
    \log_{10} \left( f_\text{esc, \lya} \right) = a \Delta v_\text{red peak, \lya} + b
\end{equation}
where $a = -0.004_{-0.002}^{+0.002}$ and $b = -0.20_{-0.39}^{+0.40}$. For RCS0224z5 specifically, the \lya\ velocity offset would result in a corresponding escape fraction of $f_\text{esc, \lya} = 0.17_{-0.07}^{+0.11}$. Since absorption by the intervening IGM could bias the \lya\ peak offset redwards, we note its measured velocity offset (and hence the implied \lya\ escape fraction) should be considered as an upper (lower) limit. Similarly, the measured velocity offset of \MgII\ implies a relatively high escape fraction. Even though the offset from the systemic redshift measured with X-shooter is somewhat uncertain (see \cref{sec:Results}), \MgII\ is $\ssim 100 \, \mathrm{km/s}$ bluewards of \lya, and $\ssim 40 \, \mathrm{km/s}$ bluewards of \CIV\ (all measured on the same X-shooter spectrum), which suggests very little scattering (resulting in P-Cygni profiles) can have taken place, indicating that both \lya\ and \MgII\ escape easily.

Independently, we can estimate the $\MgII \, \lambda \, 2796 \, \Angstrom$ escape fraction. \citet{2018ApJ...855...96H} have demonstrated a tight sequence between the \OIIIf/\OII\ ratio and the intrinsic line flux of $\MgII \, \lambda \, 2796 \, \Angstrom$ relative to $\OIIIf \, \lambda \, 5008 \, \Angstrom$. Although \OIIIf\ has not been observed directly, we can infer its flux (and the \OIIIf/\OII\ ratio) through the observed \NeIII\ and \OII\ lines, as discussed in \cref{sssec:NeIII/OII as a proxy for the ionisation state of the ISM}.

We start with \cref{eq:NeIII-OII vs OIII/OII correlation}, from which we derive an \OIIIf/\OII\ ratio of $6.6 \pm 1.6$ and hence $F_\OIIIf = (8.6 \pm 2.1) \cdot 10^{-16} \, \mathrm{erg \, s^{-1} \, cm^{-2}}$ for RCS0224z5. Again using \cref{eq:NeIII-OII vs OIII/OII correlation}, we can rewrite Equation (1) in \citet{2018ApJ...855...96H} in terms of the \NeIII/\OII\ ratio:
\begin{align}
    \label{eq:NeIII/OII R_2796 diagnostic}
    R_{2796} & = \log_{10} \left( \frac{\MgII \, \lambda \, 2796 \, \Angstrom}{\OIIIf \, \lambda \, 5008 \, \Angstrom} \right) \nonumber
    \\
    & = 0.079 \omega^2 - 1.04 \omega - 0.54 \nonumber
    \\
    & = 0.0964 \nu^2 - 0.941 \nu - 1.67,
\end{align}

\noindent where
\begin{equation*}
    \omega = \log_{10} \left( \frac{\OIIIf}{\OII} \right), \nu = \log_{10} \left( \frac{\NeIII}{\OII} \right).
\end{equation*}

\noindent The first two lines of \cref{eq:NeIII/OII R_2796 diagnostic} give the definition of $R_{2796}$ and its fitted dependence on the \OIIIf/\OII\ ratio given in Equations (1) and (3) in \citet{2018ApJ...855...96H}. The final line follows from inserting \cref{eq:NeIII-OII vs OIII/OII correlation} in this work.

For RCS0224z5, we find $R_{2796} = -1.34 \pm 0.22$, taking into account a $\ssim 0.2 \, \mathrm{dex}$ scatter in \cref{eq:NeIII/OII R_2796 diagnostic} \citep[see][]{2018ApJ...855...96H}, which translates to a predicted intrinsic $\MgII \, \lambda \, 2796 \, \Angstrom$ flux of $(5.0 \pm 1.5) \cdot 10^{-18} \, \mathrm{erg \, s^{-1} \, cm^{-2}}$. The predicted escape fraction is then $f_{\text{esc, }\MgII \, \lambda \, 2796 \, \Angstrom} = 0.13 \pm 0.08$, as indicated by the horizontal dashed line in \cref{fig:Lya/LyC-MgII escape and velocity offsets}. We furthermore note $R_{2796}$ can also be translated to the ratio of $\MgII \, \lambda \, 2796 \, \Angstrom$ over \OII, $R'_{2796}$ (which however will likely have a larger intrinsic scatter):
\begin{align}
    R'_{2796} & = R_{2796} + \omega = \log_{10} \left( \frac{\MgII \, \lambda \, 2796 \, \Angstrom}{\OII \, \lambda \, 3727, 3730 \, \Angstrom} \right) \nonumber
    \\
    & = 0.079 \omega^2 - 0.04 \omega - 0.54 \nonumber
    \\
    & = 0.0964 \nu^2 - 0.164 \nu - 0.476.
\end{align}

\noindent Finally, we will briefly discuss the feasibility of observing \MgII\ in EoR sources. Simulations\footnote{\textit{JWST} Exposure Time Calculator: \url{https://jwst.etc.stsci.edu}.} of the near-infrared spectrograph on \textit{JWST} (NIRSpec), point out that for an (intrinsically) relatively faint object like RCS0224z5, whose UV continuum would be observed as $m_\text{UV} \sim 27.4 \, \mathrm{mag}$ at $z = 7$, detecting \MgII\ spectroscopically would be challenging (see \cref{ap:JWST ETC calculation} for a detailed description). Low-resolution ($R \sim 100$) \textit{JWST}/NIRSpec observations will not resolve the doublet or provide velocity offset information. Only a very deep ($\ssim 10 \, \mathrm{h}$) exposure with \textit{JWST}/NIRSpec at medium spectral resolution would likely yield a significant detection at sufficient spectral resolution to resolve the doublet for typical $z \sim 7$ EoR galaxies, unless the observed \MgII\ flux is substantially enhanced (either intrinsically, e.g. by a luminous star-bursting galaxy, or externally by gravitational lensing). In addition, spectroscopic observations of \MgII\ could be performed with Extremely Large Telescopes out to redshift $z \lesssim 7$ in the $K$-band, in order to unlock the potential of \MgII\ not only for spectroscopic redshift confirmation, but also as an indirect tracer of \lya\ and LyC properties of galaxies in the EoR.

\section{Summary}
\label{sec:Summary}

We have presented new X-shooter and SINFONI observations of a $30 \times$ magnified galaxy at $z \simeq 4.88$, RCS0224z5. Only three sources at $z \gtrsim 5$ are known with this lensing magnification, and at the same time its redshift places RCS0224z5 just below the observational limit for accessing the bluest rest-frame optical emission lines from the ground (\OII\ and \NeIII, other lines have already shifted into the mid-infrared). This particular source has been shown to exhibit widespread, high equivalent width $\CIV \, \lambda \, 1549 \, \Angstrom$ emission, suggesting it is a unique example of a metal-poor galaxy, with a hard radiation field and high LyC production efficiency $\xi_\text{ion}$, likely representing the galaxy population that is responsible for cosmic reionisation. By virtue of its lensing magnification and redshift, RCS0224z5 is thus a unique ``Rosetta Stone'' object that could help bridge the gap in our understanding between galaxies in the local and very early Universe. We summarise our findings as follows:

\begin{itemize}
    \item We rule out the presence of strong AGN activity in the source, based on UV \citetalias{1981PASP...93....5B}-like diagnostics. Instead, a young ($1$-$3 \, \mathrm{Myr}$), metal-poor stellar population is likely responsible for the hard radiation field required for the \CIV\ emission, providing a considerable contribution of photons reaching energies of at least $47.9 \, \mathrm{eV}$.
    
    \item We present the detection of \NeIII\ (the highest redshift this line has been observed at) and discuss the potential of the \NeIII/\OII\ line ratio as a high-redshift diagnostic for the ionisation parameter (and metallicity, albeit indirect). The measured ratio of RCS0224z5, $\NeIII/\OII = 0.46 \pm 0.10$, corresponding to a likely ionisation parameter of $\log_{10} U = -2.37 \pm 0.08$, is similar to local galaxies that have been confirmed to be leaking LyC radiation, and about an order of magnitude higher than local star-forming galaxies.
    
    \item When using \NeIII/\OII\ as metallicity tracer -- indirectly, since the line ratio principally correlates with the ionisation parameter -- we estimate that RCS0224z5 has a gas metallicity of roughly $20\%$ of the solar value ($12 + \log \left ( \text{O/H} \right) = 8.01_{-0.21}^{+0.21}$), which is in mild tension with what is expected from the Fundamental Metallicity Relation (FMR). The only other galaxy known at a similar redshift ($z \sim 5$) for which \NeIII\ is detected shows a similar displacement in metallicity, as do nearby LyC leakers. Since the FMR is considered a relation describing the smooth chemical evolution of galaxies, in which galaxies are in near-equilibrium between star formation and gas inflow and outflow, the deviation of these two galaxies at $z \sim 5$ suggests that primeval galaxies (and rare, LyC-leaking galaxies in the local Universe) might be out of equilibrium by being subject to an excess of gas accretion, resulting in an excess of metallicity dilution. However, more measurements at this cosmic epoch are certainly needed to verify this trend.
    
    \item Our detection of \MgII\ in emission, the highest EW emission line observed in this source after \lya\ (and the highest redshift detection to date), demonstrates its potential for several applications. Firstly, it can simply act as a spectroscopic redshift confirmation at high redshift -- especially in the EoR (at $z \gtrsim 6$), where \lya\ will be absorbed by the neutral IGM. Secondly, since the escape of \MgII\ correlates with that of \lya\ \citep{2018ApJ...855...96H}, it might provide a new way to indirectly but effectively identify leaking LyC radiation in the same sources during an epoch when the Universe is opaque also to LyC photons. Finally, detecting \MgII\ in these sources would provide a tool to predict the intrinsic properties of \lya\ within galaxies, allowing improved constraints on the neutral fraction in the IGM derived from the \lya\ prevalence.
\end{itemize}

\section*{Data availability}

The X-shooter and SINFONI data underlying this article are available in the ESO archive at \url{https://archive.eso.org/} under ESO programme IDs 0102.A-0704(A) and 075.B-0636(B), respectively. The \textit{HST} data underlying this article are available in the MAST archive at \href{https://dx.doi.org/10.17909/t9-9kg5-hg27}{10.17909/T9-9KG5-HG27}. The reduced data underlying this article will be shared on reasonable request to the corresponding author.

\section*{Acknowledgements}

We thank Joanna Piotrowska for assisting with the dust correction of SDSS galaxy line fluxes. We are furthermore grateful to the anonymous referee for their helpful suggestions. JW, RS, R. Maiolino, and MC acknowledge support from the ERC Advanced Grant 695671, ``QUENCH'' and the Science and Technology Facilities Council (STFC). JW, RS, and R. Maiolino acknowledge support from the Fondation MERAC. RS acknowledges support from an NWO Rubicon grant, project number 680-50-1518, and an STFC Ernest Rutherford Fellowship (ST/S004831/1). NL acknowledges support from the Kavli Foundation. R. Massey and MS acknowledge funding from STFC via awards ST/T000244/1 and ST/T002565/1. This work has also used the following packages in \program{python}: the \program{SciPy} library \citep{Jones2001}, its packages \program{NumPy} \citep{2011CSE....13b..22V} and \program{Matplotlib} \citep{Hunter2007}, and the \program{Astropy} package \citep{2013A&A...558A..33A, 2018AJ....156..123A}.

Based on observations collected at the European Southern Observatory under ESO programmes 0102.A-0704(A) and 075.B-0636(B). This work was furthermore partially based on observations made with the NASA/ESA \textit{Hubble Space Telescope}, obtained at the Space Telescope Science Institute, which is operated by the Association of Universities for Research in Astronomy, Inc., under NASA contract NAS 5-26555. These observations are associated with program \#14497. Funding for the SDSS and SDSS-II has been provided by the Alfred P. Sloan Foundation, the Participating Institutions, the National Science Foundation, the U.S. Department of Energy, the National Aeronautics and Space Administration, the Japanese Monbukagakusho, the Max Planck Society, and the Higher Education Funding Council for England (\url{https://www.sdss.org/}). We are furthermore grateful to the MPA-JHU group for making their catalogue public. The MPA-JHU team was made up of Stephane Charlot, Guinevere Kauffmann, Simon White, Tim Heckman, Christy Tremonti and Jarle Brinchmann.

\bibliographystyle{mnras}
\bibliography{RCS0224z5}

\appendix

\section{\texorpdfstring{\MgII}{MgII} and \texorpdfstring{\CIII}{CIII} significance}
\label{ap:MgII and CIII significance}

\begin{table}
    \centering
    \caption[Significance of the \MgII\ detection]
    {Measured velocity offset and line flux in different subsets of the X-shooter data, corresponding to the rows in \cref{fig:MgII significance}. Given quantities are defined as in \cref{tab:Results}.
    }
    \begin{tabular}{lcc}
        \hline
        Configuration & $\Delta v \, (\mathrm{km/s})$ & $\mathrm{Flux} \, (10^{-18} \, \mathrm{erg \, s^{-1} \, cm^{-2}})$
        \\
        \hline
        \begin{filecontents*}{MgII_appendix.csv}
ptype|dv|dverr|uplim|flux|fluxerr
OB1|45|56|False|5.33|1.45
OB2|54|53|False|6.17|1.41
OB3|79|56|False|5.97|3.01
Default aperture|56|54|False|4.65|0.90
Extended aperture|52|54|False|5.00|1.49
\end{filecontents*}
\csvreader[separator=pipe, late after line=\\, head to column names]{MgII_appendix.csv}{}{\ptype & \ifcsvstrcmp{\dv}{nan}{\dots}{$\dv \ifcsvstrcmp{\dverr}{nan}{}{\pm \dverr}$} & \ifcsvstrcmp{\flux}{nan}{\dots}{\ifcsvstrcmp{\uplim}{True}{$<\flux$}{$\flux \pm \fluxerr$}}
        }
    \end{tabular}
    \label{tab:MgII significance}
\end{table}

\begin{figure*}
	\centering
	\includegraphics[width=\linewidth]{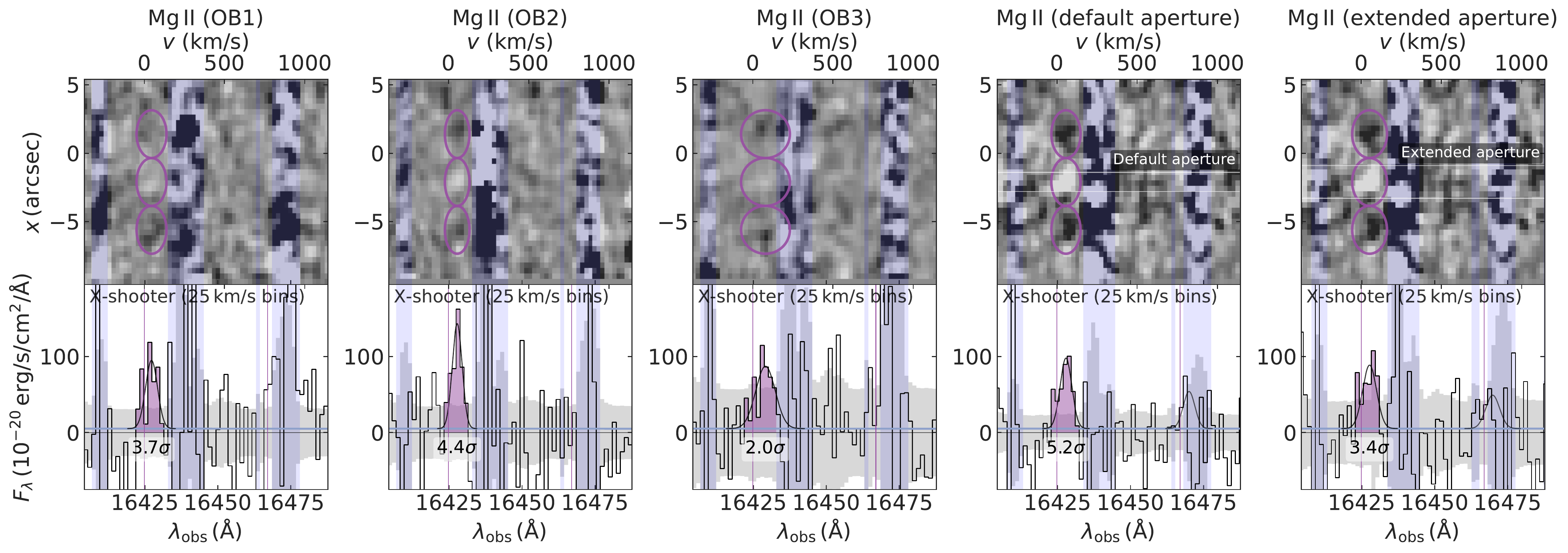}
	\caption[Various X-shooter spectra of \MgII.]{X-shooter spectra of \MgII\ for each of the three OBs individually (first three columns) and the combined spectra for a smaller and extended aperture (final two columns). One-dimensional spectra for the individual OBs have been extracted from the same smaller aperture as in the fourth column.
	}
	\label{fig:MgII significance}
\end{figure*}

In this appendix, we elaborate on the significance of the (non-)detections of the \MgII\ emission line and the $\CIIIs \, \lambda \, 1907 \, \Angstrom, \CIIIf \, \lambda \, 1909 \, \Angstrom$ doublet. In \cref{fig:MgII significance}, X-shooter spectra of \MgII\ are shown for each of the three observation blocks (OBs) individually (first three columns) and the combined result for a smaller and extended aperture (final two columns). The measured velocity offset and flux for each different configuration are summarised in \cref{tab:MgII significance}.

\begin{figure}
	\centering
	\includegraphics[width=\linewidth]{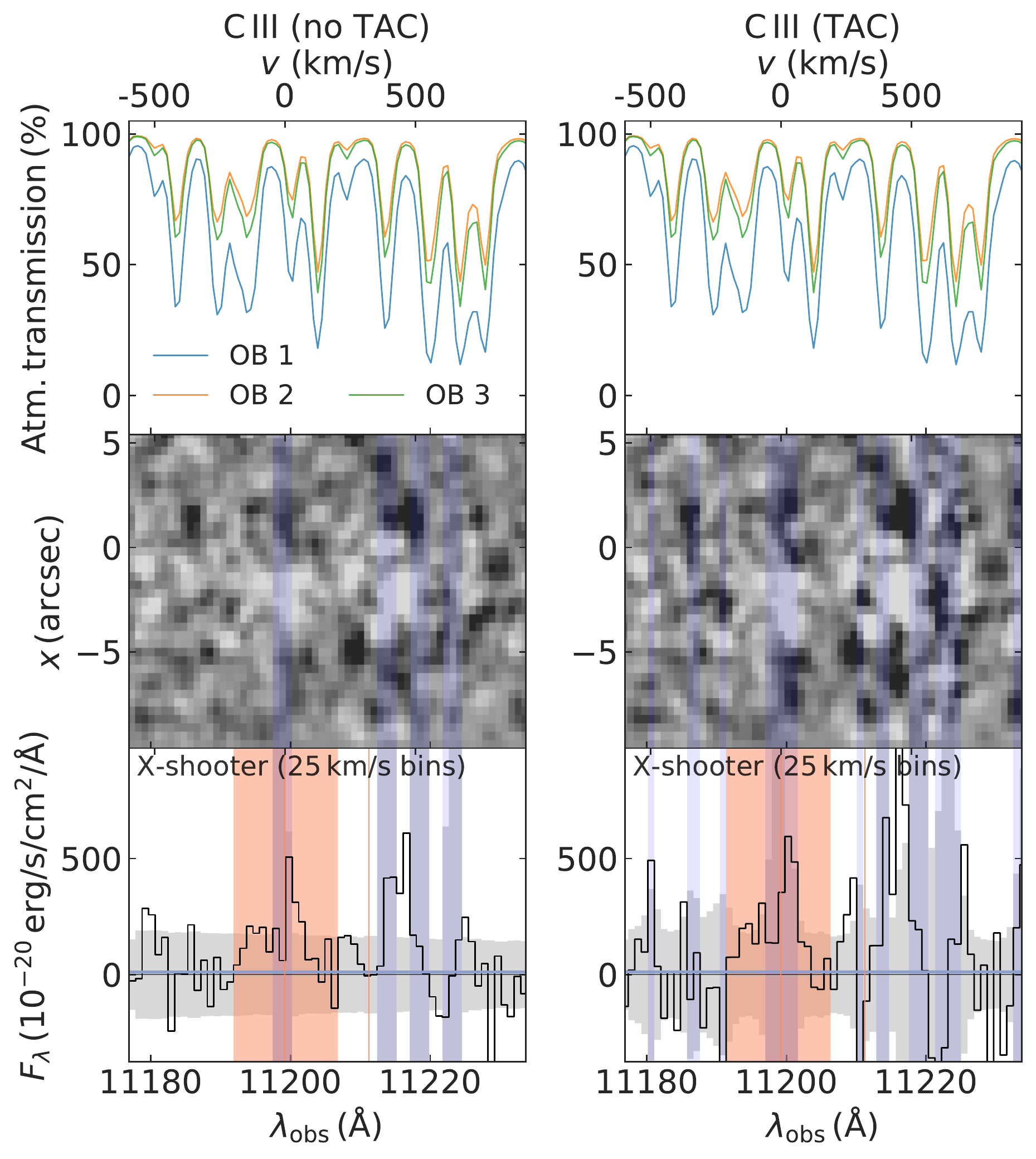}
	\caption[Non-detections of \CIII.]{X-shooter spectra in the wavelength region where the \CIII\ doublet would be expected, both without and with TAC (see \cref{ssec:Observations: X-shooter}). The top row shows the resulting atmospheric transmission calculated by \program{molecfit}. A region within $-200 \,\mathrm{km/s} < v < 200 \, \mathrm{km/s}$ of the expected $1907 \, \Angstrom$ line centre, which has been used to place an upper limit on the flux, is highlighted in the bottom row of one-dimensional spectra.
	}
	\label{fig:CIII non-detection}
\end{figure}

Furthermore, \cref{fig:CIII non-detection} shows the portion of the spectrum where the \CIII\ doublet would be expected, both without and with telluric absorption correction (TAC; see \cref{ssec:Observations: X-shooter}). It is unclear whether a signal is present in the spectra, which lack a clear dark-light-dark pattern (cf. \cref{fig:Overview panel}), in part because of skyline contamination and partly owing to the strong telluric absorption. We have chosen not to attempt to measure an upper limit for the $\CIIIf \, \lambda \, 1909 \, \Angstrom$ line directly, as it falls precisely on a region that is heavily impacted by skylines and telluric absorption. Instead, we assume a line ratio (see \cref{ssec:Results: X-shooter}).

\section{SDSS selection}
\label{ap:SDSS selection}

For the comparison sample drawn from the SDSS DR7 (discussed in \cref{ssec:Discussion: NeIII/OII}), we outline the selection criteria here in detail. Following previous studies \citep[e.g.][]{2006MNRAS.372..961K, 2014ApJ...788...88J, 2016MNRAS.456.3354F}, we select galaxies satisfying the following criteria:
\begin{enumerate}[label=(\roman*)]
    \item $\mathtt{TARGETTYPE} = \mathrm{GALAXY}$ and $\mathtt{Z\_WARNING} = 0$.
    \item For all emission lines in the ratios \OIIIf/\Hb, \NII/\Ha, \SII/\Ha, and \OI/\Ha\ used in \citetalias{1981PASP...93....5B} diagrams, we require a SNR of $\text{SNR} > 3/\sqrt{2} \approx 2.12$ on the ratios themselves (leading to a more complete sample, see \citet{2014ApJ...788...88J} -- additionally, formal uncertainty corrections as discussed in their Appendix A have been applied). Furthermore, we only select galaxies with $\text{SNR} > 30$ on the \OII\ doublet -- see the discussion in \cref{ssec:Discussion: NeIII/OII}.
    \item In order to align with previous studies, redshifts between $0.04 < z < 0.2$. These lower and upper limits are imposed to avoid strong fiber-aperture effects, and to cover detections of intrinsically weak lines while maintaining a good completeness for Seyfert-type galaxies, respectively \citep[e.g.][]{2014ApJ...788...88J}.
    \item A valid stellar mass measurement ($17$ entries have $M_* = -1$).
\end{enumerate}
This leads to a final sample of $8960$ galaxies. We classify the galaxies into star-forming, composite, Seyfert, and LINER classes (although we will focus only on star-forming and Seyfert types), based on the \NII, \SII, and \OI\ \citetalias{1981PASP...93....5B} diagrams, following \citet{2006MNRAS.372..961K}. Subsequently, the line fluxes are corrected for dust extinction using the \citet{1989ApJ...345..245C} reddening curve assuming $R_V = A_V/E(B-V) = 3.1$, and a fiducial intrinsic \Ha/\Hb\ ratio of $2.85$ for star-forming galaxies, and $3.1$ for AGN-dominated systems \citep[for case-B recombination at $T = 10^4 \, \mathrm{K}$ and $n_e \sim 10^2$-$10^4 \, \mathrm{cm^{-3}}$, see][]{2006MNRAS.372..961K}. In this sample, $2484$ galaxies or 27.7\% have a $\text{SNR} > 5$ \NeIII\ detection.

\section{\textit{JWST} ETC calculation}
\label{ap:JWST ETC calculation}

Given the observed $\MgII \, \lambda \, 2796 \, \Angstrom$ flux of $5.0 \cdot 10^{-18} \, \mathrm{erg \, s^{-1} \, cm^{-2}}$ (see \cref{tab:Results}) and assuming a typical flux ratio of $F_{2796}/F_{2804} \approx 1.9$ between the $\MgII$ lines at $2796 \, \Angstrom$ and $2804 \, \Angstrom$ \citep[e.g.][]{2018ApJ...855...96H}, the total flux of the doublet would become $7.6 \cdot 10^{-18} \, \mathrm{erg \, s^{-1} \, cm^{-2}}$. However, taking into account the lensing magnification of $\mu = 29$, we derive an intrinsic flux of $2.6 \cdot 10^{-19} \, \mathrm{erg \, s^{-1} \, cm^{-2}}$ at $z=4.88$. We note that the uncertainty and spatial variation of the lensing magnification makes this only a rough estimate of the true intrinsic flux. Assuming an object with the same luminosity at $z = 7$ (in which case \MgII\ would be observed at $\lambda_\text{obs} = 2.24 \, \mathrm{\upmu m}$), this would lead to an observed flux of $1.13 \cdot 10^{-19} \, \mathrm{erg \, s^{-1} \, cm^{-2}}$. The continuum flux density from our fit is $2.29 \cdot 10^{-20} \, \mathrm{erg \, s^{-1} \, cm^{-2} \, \Angstrom^{-1}}$ or $383 \, \mathrm{nJy}$ at $\lambda_\text{obs} = 2.24 \, \mathrm{\upmu m}$, which translates to $2.5 \cdot 10^{-22} \, \mathrm{erg \, s^{-1} \, cm^{-2} \, \Angstrom^{-1}}$ or $4.18 \, \mathrm{nJy}$ if it were unlensed at $z=7$.

Alternatively, our estimate implies $F_{2796} = 7.4 \cdot 10^{-20} \, \mathrm{erg \, s^{-1} \, cm^{-2}}$. This is inconsistent with a recent estimate from \citet{2020MNRAS.498.2554C}, the difference being explained by the fact that their higher flux estimate (by a factor $\ssim 8$) arises from considering a source with a $H$-band magnitude of $25$ (in the F160W filter; this corresponds to $M_\mathrm{UV} \simeq -21.9$ at $z=7$). The unlensed observed magnitude of RCS0224z5 is $\ssim 26.8 \, \mathrm{mag}$, which at $z \simeq 4.88$ translates to $M_\mathrm{UV} \simeq -19.6$. This implies an observed magnitude of $27.4$ at $z=7$ -- a factor $\ssim 9$ fainter than a $25 \, \mathrm{mag}$ source -- and is instead appropriate when considering intrinsically fainter and hence more common sources. In a typical extremely deep field, one would on average expect less than one source at $z \sim 7$ with magnitude $25$, and the order of $N \sim 1$ in a medium-deep field, compared to $N \sim 2$ and $N \gtrsim 70$ respectively for a $m_\text{UV} \sim 27.4 \, \mathrm{mag}$ source (derived from the $z \sim 7$ number counts in the $5 \, \mathrm{arcmin^2}$ XDF and $\ssim 120 \, \mathrm{arcmin^2}$ CANDELS-DEEP fields presented in \citealt{2015ApJ...803...34B}).

Simulations\footnote{\textit{JWST} Exposure Time Calculator: \url{https://jwst.etc.stsci.edu}.} of the near-infrared spectrograph on \textit{JWST} (NIRSpec), point out that for an (intrinsically) relatively faint object like RCS0224z5, detecting \MgII\ spectroscopically would be challenging: for example, a $10 \, \mathrm{ks}$ exposure with the multi-object spectrograph observing mode at low resolution would result in a signal of \MgII\ at the level of $0.65 \sigma$ per spectral pixel (but we note that integrating over the few pixels containing the line could slightly increase the overall SNR). At $R \sim 100$, this would render the doublet (separated by $\ssim 770 \, \mathrm{km/s}$) unresolved as well. The same exposure would yield a SNR of $0.56$ per spectral pixel at medium spectral resolution (the F170LP/G235M grating achieving $R \sim 1000$ at $\lambda_\text{obs} = 2.24 \, \mathrm{\upmu m}$), which would resolve the doublet down to $\ssim 300 \, \mathrm{km/s}$. Finally, a deep ($10 \, \mathrm{h}$) exposure allow a SNR of $0.67$ per spectral pixel at high resolution ($R \sim 2000$ or $\ssim 150 \, \mathrm{km/s}$). Still, for objects with a more intense episode of ongoing star formation (possibly boosting the flux by a factor of a few, up to a factor $\ssim 8$ for a $M_\mathrm{UV} \simeq -21.9$ source as in \citealt{2020MNRAS.498.2554C}, as discussed above), or for lensed objects (like RCS0224z5), observations of \MgII\ would be feasible in deep spectroscopic surveys.


\bsp	
\label{lastpage}
\end{document}